\documentclass[12pt,preprint]{aastex}

\shorttitle{Extinction law and dust excitation in NGC\,300}
\shortauthors{Roussel et al.}

\begin{document}

\title{Extinction law variations and dust excitation in the spiral galaxy NGC\,300}

\author{H. Roussel\altaffilmark{1},
A. Gil de Paz\altaffilmark{2},
M. Seibert\altaffilmark{1},
G. Helou\altaffilmark{1},
B.F. Madore\altaffilmark{2},
C. Martin\altaffilmark{1}
}

\affil{1) California Institute of Technology, Pasadena, CA 91125}
\affil{2) Carnegie Observatories, Carnegie Institution of Washington, Pasadena, CA 91101}
\email{hroussel@irastro.caltech.edu}

\begin {abstract}
We investigate the possible origin of the strong radial gradient in
the ultraviolet-to-infrared ratio in the nearby late-type spiral galaxy
NGC\,300, and emphasize the importance of local variations in the geometry
of the interstellar medium,
concluding that they cannot be neglected with
respect to metallicity effects.
This analysis is based upon a combination
of maps obtained both from space using {\small GALEX} (1350 to 2750\,\AA)
and Spitzer (3 to 160\,$\mu$m), and from the ground (UBVRI, H$\alpha$ and
H$\beta$). We select ionizing stellar clusters associated with \ion{H}{2}
regions of widely varying morphologies, and derive their fundamental
parameters from population synthesis fitting of their spectral energy
distributions, carefully measured to eliminate local background emission
accurately.
From these fits, we conclude that the stellar extinction law is highly
variable in the line of sight of young clusters of similar ages.
In the particular model geometry that we consider most appropriate to the
sampled regions, we checked that our findings are not significantly altered
by the correct treatment of radiative transfer effects.
The variations are systematic
in nature, in the sense that extinction laws of the Milky Way or LMC type,
with a prominent 2175\,\AA\ absorption feature and a modest far-ultraviolet
slope, are associated with compact \ion{H}{2} regions (the compacity being
quantified in two different ways), while clusters surrounded by diffuse
\ion{H}{2} regions follow extinction laws of the 30\,Doradus or SMC type,
with a weaker 2175\,\AA\ feature and a steeper far-ultraviolet rise.
The Calzetti starburst attenuation law, although most often degenerate
with the 30\,Doradus extinction law, overpredicts ionizing photon fluxes
by large amounts. We also find that the extinction law variations are
correlated with the column density of dust species emitting in the near-
and mid-infrared. Finally, we briefly discuss the nebular to stellar
extinction ratios, and the excitation of aromatic band carriers, invalidating
their claimed association with cold dust.
\end {abstract}
 

\section{Introduction}

NGC\,300 is a gas-rich galaxy of Sd type, at a distance of only 2.1\,Mpc
\citep{Freedman}. Its total extent in the blue light is over 20\arcmin,
and its inclination on the line of sight, derived from \ion{H}{1} kinematics,
is about 50\degr\ \citep{Puche}. It contains numerous bright \ion{H}{2}
complexes \citep{Sersic, Deharveng} and forms stars at a current total
rate of about 0.15\,M$_{\sun}$\,yr$^{-1}$ \citep{Helou04}. NGC\,300 was
targetted for surveys of supernova remnants~; between 20 and 40 candidates
were found by various selection methods \citep{Blair, Pannuti, Payne},
evidence that star formation has been going on for an extended period at
a significant rate. Large numbers of Wolf-Rayet stars were also found
\citep{Breysacher, Schild}. The oxygen abundance in the nebular gas phase
varies between about 1.4 to 0.4 times the solar neighbourhood value, from
the center up to a radial distance of about 10\arcmin\ \citep{Deharveng}.

Many bubble-like structures are seen in the H$\alpha$ emission, especially in
the outer disk, where they reach diameters of $\sim 600$\,pc. About
50\% of the total H$\alpha$ flux arises from the diffuse ionized gas component
\citep{Hoopes}. Judging from the morphology of \ion{H}{2} regions and
the position of the most likely exciting sources, the hydrogen gas may be
ionized very far away from young clusters (in projected distance).
This argues for the existence of aged star-forming sites and a high degree
of porosity of the interstellar medium around them. The large number of
resolved stellar clusters and the variety of their associated \ion{H}{2}
regions, observed at a resolution of a few tens of parsecs, make it
possible to study in great detail general patterns in the extinction
and dust heating within a single galaxy.

Deriving spatially-resolved extinction in galaxies of different types,
and determining how the extinction law is affected by properties of the
interstellar medium such as metallicity and geometric configuration,
is a first step toward assessing the validity of common recipes used to
correct for extinction the global ultraviolet emission of galaxies. This
knowledge is fundamental to obtaining good prescriptions to estimate star
formation rates and ages from current and future ultraviolet and infrared
surveys.

In this paper, we analyse jointly images of the stellar light from the
far-ultraviolet to the far-red, of the hydrogen recombination emission in the
H$\alpha$ and H$\beta$ lines, and of the infrared emission from 3 to 160\,$\mu$m.
The data are described in Section~\ref{data}. After deriving large-scale
infrared-to-ultraviolet ratio and extinction maps (Section~\ref{gradient}),
we focus on individual star-forming complexes selected in H$\alpha$ and in
the ultraviolet, and constrain the fundamental properties of ionizing stellar
clusters, to assess how critically the extinction law and excitation of
mid-infrared emitting dust at these spatial scales depend on the local
geometry of the interstellar medium (Section~\ref{complexes}). The results
are discussed topic by topic in Section~\ref{discussion}, and summarized
in Section~\ref{summary}.

\section{Data}
\label{data}

\subsection{Ultraviolet images}

NGC\,300 was observed with the {\small GALEX} space telescope in the
FUV and NUV bands at effective wavelengths of 1516 and 2267\,\AA\
\citep{Martin, Morrissey}. {\small GALEX} has a circular field of view
of 1.2\degr\ in diameter, and observations are performed in a spiral
dither pattern. The data from the photon-counting, multichannel-plate
detectors were processed with the analysis pipeline operated by the
Caltech Science Operations Center, which converts the photon lists into
flux-intensity images and applies flat-fielding and first-order geometric
distortion corrections. The absolute calibration is derived from ground
test data which show good agreement, to within 10\%, with observations
of white dwarf standard stars \citep{Morrissey}. See also the {\small GALEX}
Atlas of galaxies \citep{Paz}, where NGC\,300 is included, for more
information.

The angular resolution is of the order of 5\arcsec\ (FWHM), but varies
with source brightness and position across the field. In particular,
the point spread function in the FUV band is slightly elliptical (elongated
along the south-east to north-west direction) in some parts of the map.
A first distortion correction between the NUV and FUV maps was
estimated using Legendre polynomials of order 4 and cross-correlating
the positions of about 900 stars and clusters. Distortion residuals
and point spread function variations were handled separately
for each individual sub-field of size $\sim 2$\arcmin, by fitting
nearby point sources with gaussians in the FUV, NUV and U bands.

\subsection{Infrared images}

Broadband infrared images were acquired with the Spitzer space telescope
at effective wavelengths of 3.6, 4.5, 5.7, 8.0, 24, 71 and 156\,$\mu$m,
and were first discussed by \citet{Helou04}. The data processing and
mosaicing is also described in this paper. The angular resolution (full
width at half maximum) of the mosaics is 2.5 to 3\arcsec\ between 3.6
and 8\,$\mu$m, 5.7\arcsec\ at 24\,$\mu$m, 16\arcsec\ at 70\,$\mu$m and
38\arcsec\ at 160\,$\mu$m.

The mid-infrared bands record gradually varying contributions, as a
function of wavelength, from stars and dust, the global 3.6\,$\mu$m
emission being dominated by photospheres, the global 8\,$\mu$m emission
deriving mainly from aromatic band carriers, and longer-wavelength bands
covering essentially the thermal continuum from very small to big dust grains.

The photospheric component was subtracted from the 8\,$\mu$m map as
explained in \citet{Helou04}. A model for intermediate to old stellar
populations was derived from the Starburst99 population synthesis model
\citep{Leitherer}, and was found to be remarkably invariant for solar
and half-solar metallicities, and a broad range of star formation histories
(excluding very young populations). Assuming that the 3.6\,$\mu$m emission
is entirely of stellar origin, this map was then used to normalize the
photospheric spectrum at each point, to which the relevant filter transmission
curves were applied. This procedure may be inaccurate locally, because
of the contribution from young stars, but the 8\,$\mu$m emission is
heavily dominated by dust in star-forming regions, so that neglecting
young stellar populations should not introduce any significant error.

\subsection{Optical images}

Mosaics in the UBVI bands of $20.5\arcmin \times 20.5$\arcmin\ were
obtained from S.C. Kim \citep[private communication]{Kim}. An R-band mosaic
of $25\arcmin \times 25$\arcmin\ and H$\alpha$+[\ion{N}{2}] and H$\beta$
maps in narrowband filters (in a field of view of
$8.85\arcmin \times 8.85$\arcmin) were acquired at the Las Campanas
du Pont 2.5\,m telescope with the direct CCD camera. The angular resolution
in all the broadband optical images, after the mosaicing and the geometric
transformations applied to align them, is of the order of 1.7\arcsec\
to 2\arcsec\ (FWHM).

We also made use of an H$\alpha$ mosaic covering the same field as the
R-band mosaic to measure the H$\alpha$ flux of the outermost \ion{H}{2}
regions~; the residual background was subtracted locally in each subfield
separately. The map acquired by \citet{Hoopes}, which was kindly provided
to us by Charles Hoopes, supplied a means of checking the H$\alpha$+[\ion{N}{2}]
flux calibration, by comparing aperture photometry measurements
on several bright \ion{H}{2} complexes in both maps~; fluxes from the map
of \citet{Hoopes} are systematically lower by about 10\%.

The H$\beta$ photometric accuracy, however, is much more uncertain,
because this line has a much smaller equivalent width than the H$\alpha$
line, and larger errors are introduced by continuum subtraction and
correction for stellar absorption features. Because of these uncertainties,
if no flux correction is applied, the H$\alpha$/H$\beta$ decrement values
are unphysical in the case B recombination assumption, and in disagreement
with published values for individual \ion{H}{2} regions.
In order to correct them and derive an estimate of the nebular extinction
from H$\alpha$ and H$\beta$ photometry, we scaled the decrement in such a
way that extinction values derived by \citet{Webster} from spectro-photometry
for 5 regions could be reproduced. We used an intrinsic H$\alpha$/H$\beta$
ratio of 2.86, appropriate for electronic densities and temperatures
$\leq 10^3$\,cm$^{-3}$ and $\sim 10^4$\,K, respectively, and replaced
the extinction law assumed by \citet{Webster} with the \citet{Cardelli}
law. This procedure requires that the raw H$\alpha$/H$\beta$ flux ratio
be underestimated by 30\% on average. Applying the above correction, the
derived global extinction (from total fluxes in the H$\beta$ field of
view) is A(H$\alpha$) = 0.5\,mag.

\section{Large-scale UV/IR ratio and extinction}
\label{gradient}

\begin{figure}[!ht]
\hspace*{-1cm}
\resizebox{18cm}{!}{\rotatebox{90}{\plotone{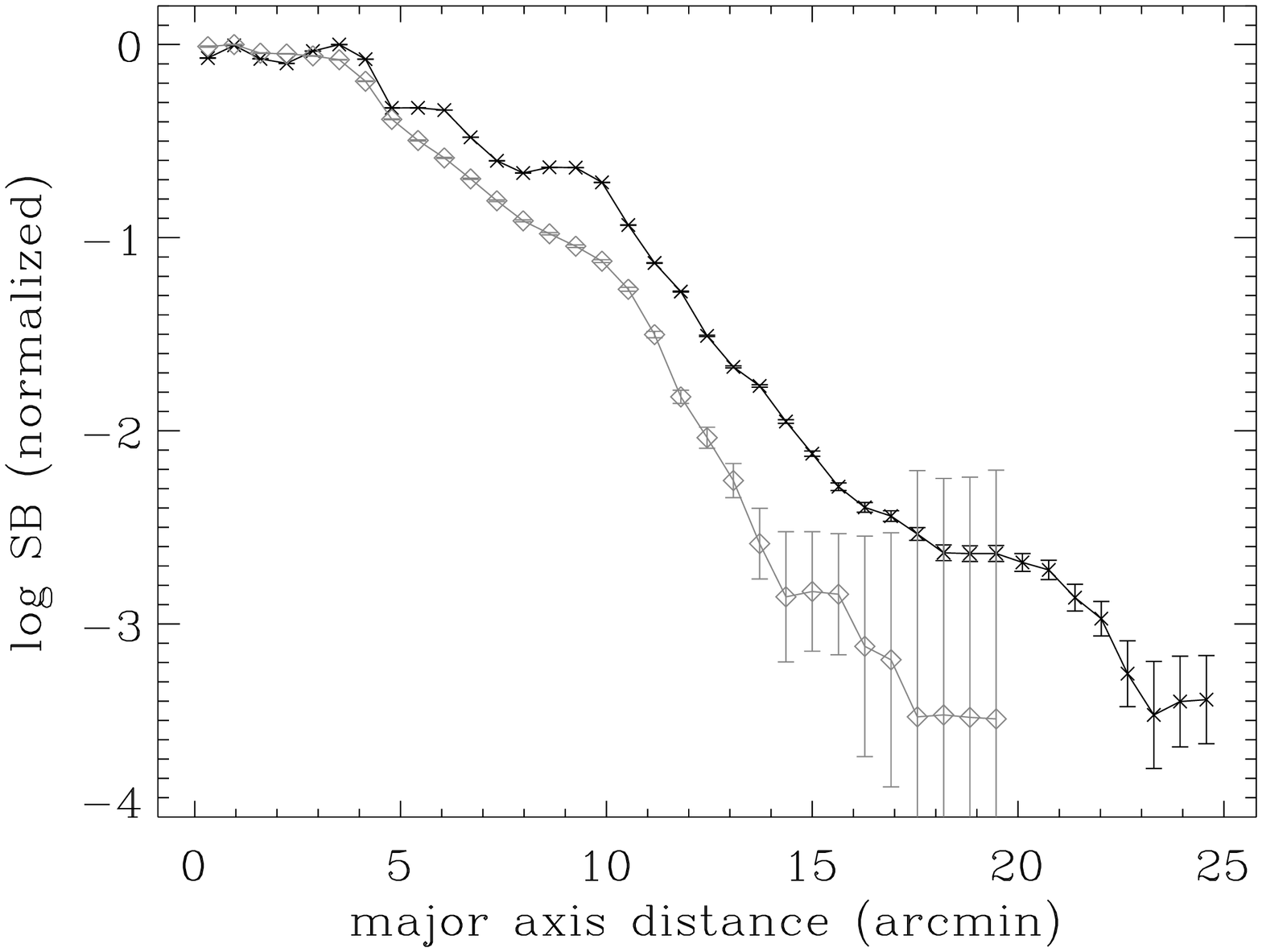}}\rotatebox{90}{\plotone{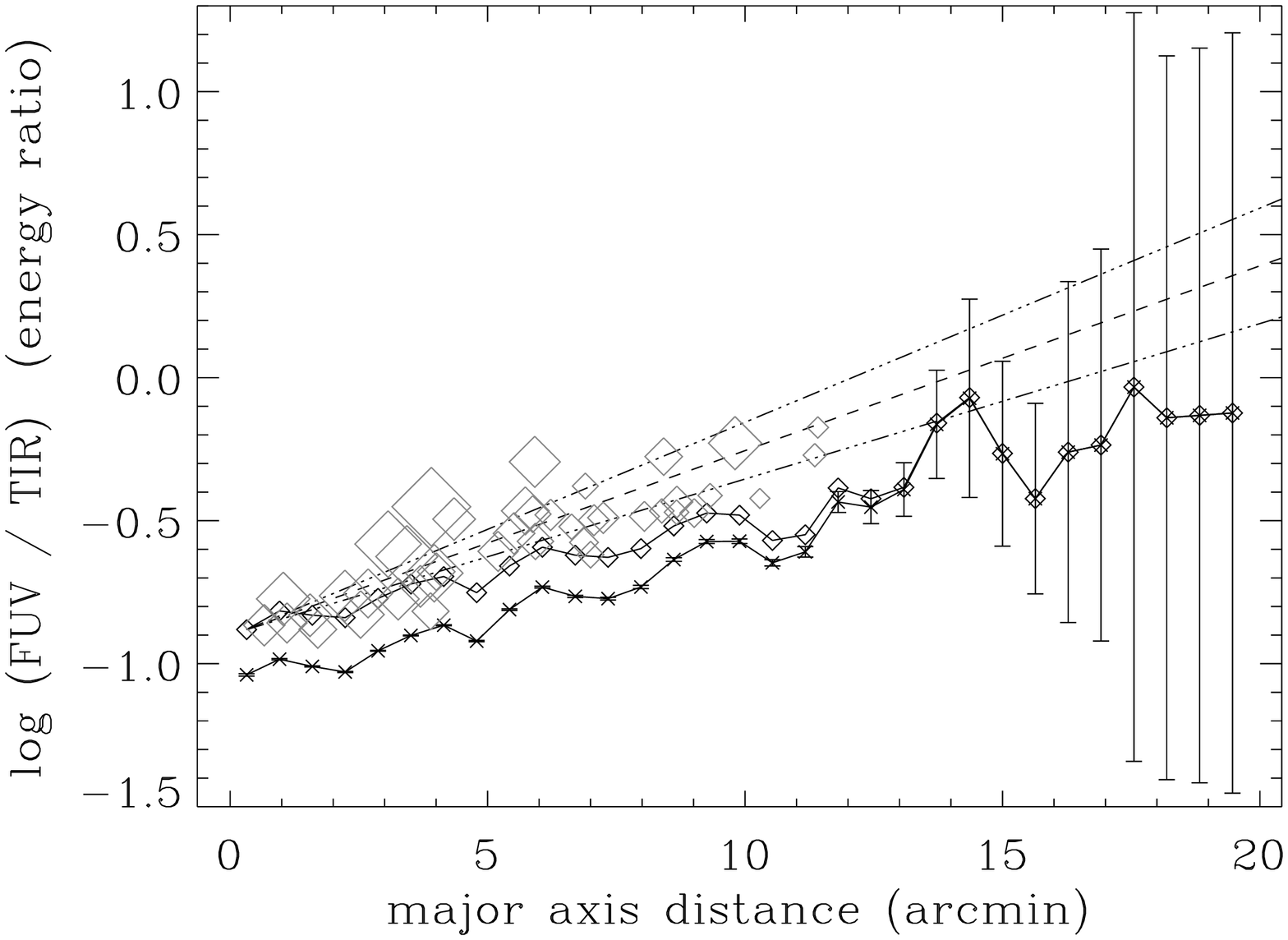}}}
\caption{{\bf (a)} Radial profiles of the FUV and infrared surface
brightnesses, measured in elliptical annulli oriented along the galaxy
major axis. The black line corresponds to the FUV band and the grey line
connecting the diamonds to the total infrared emission estimated from the
24, 70 and 160\,$\mu$m maps. The curves are normalized by their maximum value.
{\bf (b)} Radial profile of the FUV to infrared energy ratio, before
(star symbols) and after (diamond symbols) correction of the FUV map for
extinction. The energy emitted in the FUV band was computed with a filter
equivalent width of 268\,\AA. The linear dashed lines represent the
opposite of the metallicity gradient derived by \citet{Deharveng} (with its
uncertainty range). The big grey diamonds are individual FUV sources,
measured at the same angular resolution of 38\arcsec, within 76\arcsec\
apertures~; their size is proportional to the square root of the FUV flux
corrected for extinction, using the results from Figure~\ref{fig:a_fuv}.
}
\label{fig:profils}
\end{figure}

The ultraviolet disk reaches much farther out than the infrared disk.
To quantify this statement, we convolved all maps to the resolution of
the 160\,$\mu$m map, and estimated the total infrared emission (TIR) from
a linear combination of the 24, 70 and 160\,$\mu$m powers \citep{Dale02}.
To reach the 160\,$\mu$m full width at half-maximum, the maps at wavelengths
up to 24\,$\mu$m were first convolved with a model of the point spread
function at 70\,$\mu$m\footnote{which can be found on the website:
{\it http://ssc.spitzer.caltech.edu/mips/psf.html\,.}}~; these and the
70\,$\mu$m map were then convolved with gaussians of the appropriate widths.
We computed surface
brightness profiles in the FUV band and in the TIR, averaged in elliptical
annuli to account for the inclination of NGC\,300 (Fig.~\ref{fig:profils}).
The FUV band was preferred over the NUV band for this comparison, because
while the ratio of the extinction in these two bands is close to 1
(smaller than 1.3 for various extinction laws), the intrinsic FUV/NUV
ratio of young stellar populations is $\geq 1$, the FUV band avoids the
variable 2175\,\AA\ feature of the interstellar extinction curve
(see Section~\ref{extlaw}), and the FUV band is minimally contaminated
by foreground stars.
On average, the FUV/TIR ratio increases exponentially with deprojected
distance from the center, with irregularities mainly due to the FUV
emission being less smooth than the infrared emission. The FUV/TIR
ratio increases by at least a factor 5, maybe up to a factor 10,
between the center and the outermost regions where it is still measurable.
Also shown in Figure~\ref{fig:profils} is the opposite of the average slope of
the metallicity gradient as derived by \citet{Deharveng}. The slope of the
FUV/TIR gradient is very similar, although somewhat smaller. At any given
distance from the center, however, the dispersion in FUV/TIR ratios among
individual star-forming regions is at least a factor 2, whereas the standard
deviation of abundances is only of 0.11\,dex, which seems to indicate
that metallicity effects are not solely responsible for the FUV/TIR
gradient. The UV-brightest regions tend to be deficient in infrared
emission compared with their fainter counterparts.
From studies of large samples of galaxies, a clear decrease of the global
ultraviolet to infrared flux ratio was found as a function of a measure of the
star formation rate \citep{Wang}, or as a function of the blue luminosity
\citep{Buat} or the stellar mass \citep{Pierini}. Interpreting the
ultraviolet to infrared ratio variations within resolved galaxies
could help constrain the underlying causes of this behavior.

A stellar extinction map of NGC\,300 can be derived from a simple energy
balance calculation, using all the maps between 1516\,\AA\ and 3.5\,$\mu$m
to reconstruct the stellar emission. The emerging fluxes in the FUV, NUV,
UBVRI and IRAC1 (3.5\,$\mu$m) bands were interpolated by realistic stellar
spectral energy distributions, using the actual filter transmission curves.
Assuming several different extinction laws, the power absorbed in this range
is set equal to the total power emitted in the infrared.
The derived A(FUV) ranges between 0.1 and 0.6 (A(V) between 0 and 0.23).
Uncertainties of 20\% in the flux calibration at both 70 and 160\,$\mu$m
could increase the A(FUV) extinction values by 0.05\,mag.

The large \ion{H}{2} complexes including shell-like structures correspond
to regions of lower extinction, consistent with the idea that they are
aged star-forming sites having partly dispersed the molecular and dust material
around them. The correspondance with gas column density cannot be tested
because no map in a molecular transition is available. The \ion{H}{1} map
of \citet{Puche}, with a beam size of 17\arcsec, allows us to see partial
correspondance with the \ion{H}{1} column density, though, especially in
the outer disk (Fig.~\ref{fig:hi_apertures})~; the fact that there is no
robust correlation with the extinction map is expected because \ion{H}{1}
does not trace adequately the total gas mass, which is usually dominated by
the molecular phase in the inner parts of the disk where star formation
is taking place actively.

The radial profile of the ratio of the total stellar power (integrated
between 1000\,\AA\ and 3.5\,$\mu$m) to the total infrared power is notably
different from the FUV/TIR profile: it shows a sharp decline from the
center up to the pseudo-ring of most active star formation, reaching a
minimum around a radial distance of 3 to 4\arcmin, then rises again to
reach its central value at a radial distance close to 10\arcmin.

Several effects can contribute to the large-scale FUV/TIR gradient. \\
1) The most obvious possibility is the variations in metallicity,
which are expected to cause direct variations in the dust to gas ratio.
However, various lines of evidence suggest that other parameters play
a more fundamental role. Although the global dust to gas ratio of galaxies
seems correlated with the far-infrared surface density \citep{Andreani},
which traces both the column density of dust and the star formation rate
surface density, the dust to \ion{H}{1} mass ratio of irregular galaxies
is highly variable but not correlated with metallicity \citep{Hunter}.
\citet{Sauvage} also showed that the far-infrared to blue flux ratio
of dwarf galaxies is not correlated with their oxygen abundance. Within NGC\,300,
the nebular gas phase metallicity gradient, as determined by \citet{Deharveng},
is slightly steeper than the FUV/TIR gradient. Interestingly, \citet{Bot}
found that the dust to gas ratio of the Small Magellanic Cloud is much
lower than expected from a linear scaling with metallicity, although the
situation may differ between star-forming regions and the diffuse
interstellar medium. \\
2) The large-scale variations of the FUV/TIR ratio may also be affected
significantly by the cumulative effects of local variations in the
geometry of the sources. Several factors can contribute to systematically
increase the distance between stellar clusters and the absorbing dust,
notably: a decrease in the volume density of all the phases of the
interstellar medium~; the maturity of the star forming regions,
i.e. the outer disk could contain slightly more evolved young stellar
populations which have had time to disperse gas and dust away.
Aged star formation complexes indeed seem to efficiently increase
the porosity of the surrounding interstellar medium, because they are
associated with diffuse, shell-like \ion{H}{2} regions, and are
affected by lower extinction. \\
3) The spectral energy distribution of the heating radiation could
change in such a way that the total dust heating efficiency decreases.
The fraction of the total stellar power emitted in the FUV band
varies from less than 2\% in the central region of the galaxy to
about 5\% in the outermost part of the disk, at and beyond a radial
distance of 10\arcmin\ (the whole SED being corrected for extinction).
The non-ionizing part of the stellar radiation thus becomes on average
more energetic in the outer disk (the old populations have a smaller
scale length than the younger populations emitting in the ultraviolet),
and the dust heating balance should shift progressively from big grains
emitting in the far-infrared to smaller grains emitting in the mid-infrared,
which could explain part of the far-infrared flux deficit. We indeed
observe a slight increase of the $F_{24}/{\rm TIR}$ fraction in the outer
disk (by about 30-40\% compared with the annulus of intense star formation).
However, this increase is too modest to account for the FUV/TIR gradient,
which is mirrored closely by a similar ${\rm FUV}/F_{24}$ gradient.

\begin{figure}[!ht]
\resizebox{12cm}{!}{\rotatebox{90}{\plotone{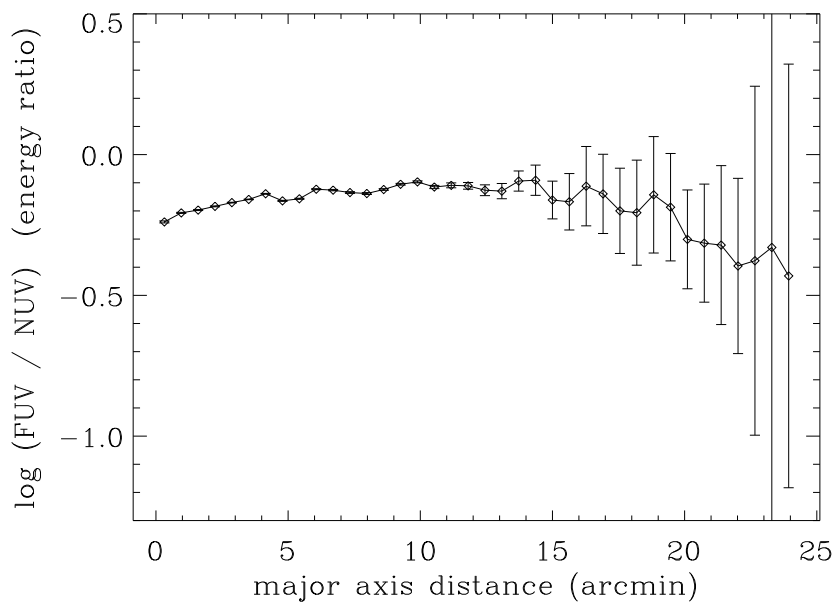}}}
\caption{Radial profile of the FUV to NUV energy ratio, computed in the
same way and at the same angular resolution as the FUV/TIR profile.
Since the differential extinction between the two bands is very small,
the extinction-corrected profile, if plotted, would be almost
indistinguishable. We used filter equivalent widths of 268\,\AA\
and 732\,\AA, respectively. The flux density ratio is higher by a
factor 1.22, or 0.087\,dex. The error bars represent the $3 \sigma$
uncertainties on the ratio.
}
\label{fig:fuv_nuv}
\end{figure}

Figure~\ref{fig:fuv_nuv} shows the radial profile of the FUV/NUV energy
ratio, which decreases on average beyond a radial distance of
$\sim 10$\arcmin. Such a decrease of the FUV/NUV ratio in the outermost
part of the disk cannot be explained by extinction or metallicity effects.
It thus seems reasonable to assume that a gradient exists in the age
of the youngest stellar populations which radiate in the ultraviolet.
The range of FUV/NUV colors observed at large scales corresponds to
average stellar population ages between about 16\,Myr at radial distances
near 10\arcmin, and $\sim 300$\,Myr in the outermost disk.

In order to assess the role played by the geometry of the interstellar
medium in the FUV/TIR variations, and to distinguish between the causes
mentioned above, we now turn to individual \ion{H}{2} complexes showing
a wide range of morphologies, presumably corresponding to very different
geometries. We take advantage of the high spatial resolution of the data
from the UV to the near-infrared, and characterize individual stellar
clusters or small groups of clusters. If each one consists of coeval
populations, then it is possible to lift the star-formation history
degeneracy that affects the large-scale emission.

\section{Small-scale properties of \ion{H}{2} complexes}
\label{complexes}

\subsection{Selection of \ion{H}{2} regions/young stellar clusters}

\begin{figure}[!ht]
\vspace*{-2.5cm}
\hspace*{-1cm}
\resizebox{17cm}{!}{\plotone{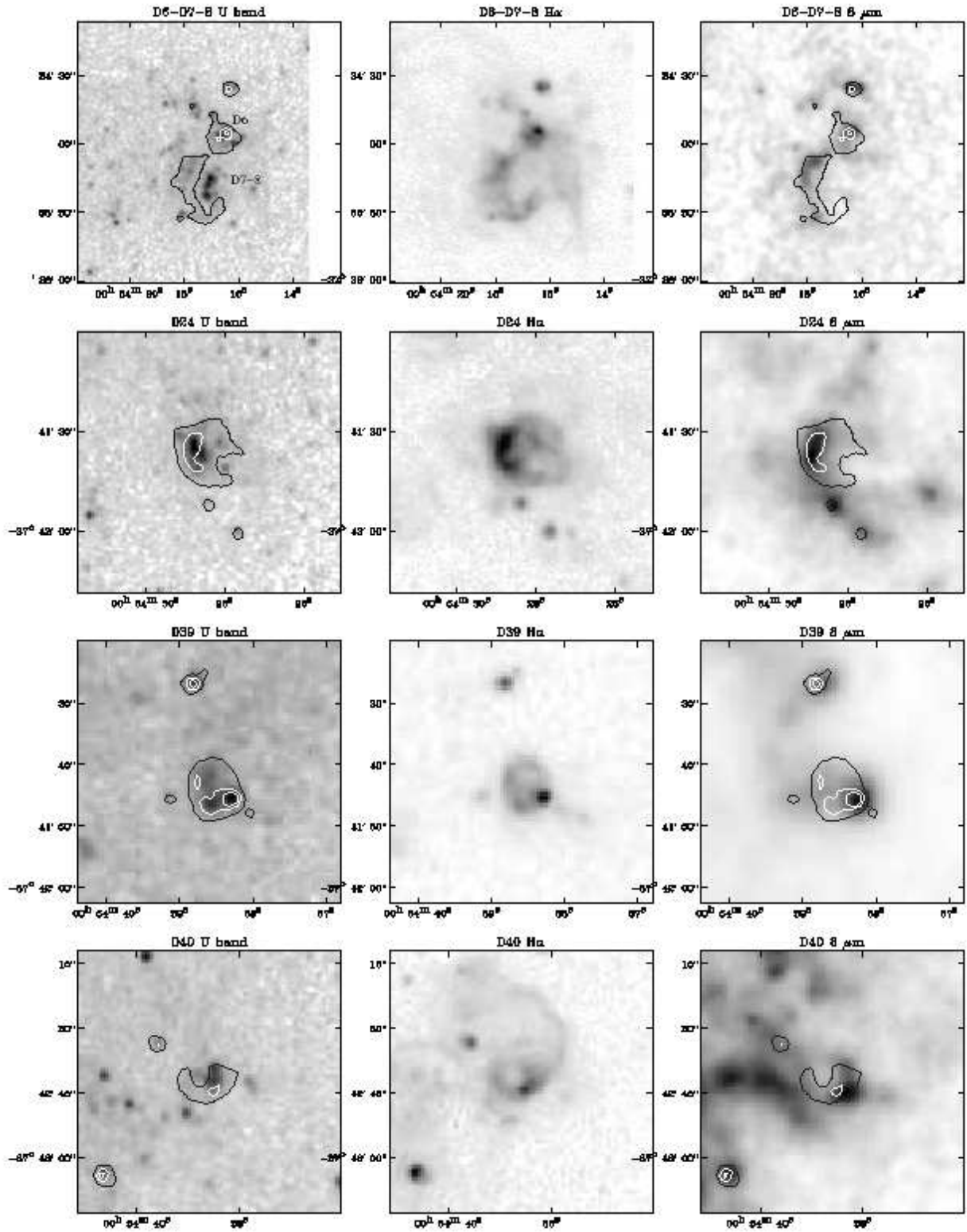}}
\caption{\footnotesize{Maps of the complexes in U (preferred to FUV and NUV because
of the more advantageous angular resolution), in H$\alpha$, and at 8\,$\mu$m.
The horizontal axis represents the right ascension and the vertical axis the
declination, in the J2000 equatorial coordinate system.
The scale varies from
one complex to another. H$\alpha$ contours are superposed onto the U-band
and 8\,$\mu$m images. The intensity scale is linear for the 8\,$\mu$m images,
square root for U and H$\alpha$. {\bf Figure abridged for astro-ph submission.}
}}
\label{fig:images}
\end{figure}

We selected individual stellar clusters or groups of clusters associated
with prominent \ion{H}{2} regions, that are visible in all the optical
bands and sufficiently bright in the FUV and NUV bands to allow accurate
photometry. 
Because of the selection method, the sample is biased against highly
obscured regions, for which the analysis described in Section~\ref{text_fits}
would not be possible. Bright 24\,$\mu$m sources which are not detected
in the ultraviolet are discussed in Section~\ref{extremes}.
We also avoided objects with very complicated structure
(making the identification of the clusters ionizing a particular
H{\small II} region ambiguous),
with contamination by foreground stars or blending with nearby regions.
The sample covers a large range of morphologies, mid-infrared colors,
and UV to infrared ratios. They were labelled according to the \ion{H}{2}
region which they excite, in the catalog of \citet{Deharveng} (but the
letter sublabels follow a different convention). Detailed notes, pertaining
both to the morphology and to the modelling results, can be found in
Appendix~\ref{annexe}. Images of the cluster fields in the U band, in
H$\alpha$ and at 8\,$\mu$m are shown in Figure~\ref{fig:images}.

Table~\ref{tab_clusters} lists, for each group of clusters, the initial
diameter of the stellar photometric aperture (see Sect.~\ref{photom})
and some observables: the H$\alpha$ flux (corrected for an average
[\ion{N}{2}] contribution to the total H$\alpha$+[\ion{N}{2}] flux of 10\%),
the nebular extinction derived from the H$\alpha$/H$\beta$ decrement,
and the H$\alpha$/R ratio measured within the stellar photometric area,
R referring to line-subtracted broadband flux (i.e. pure stellar emission).
H$\alpha$ equivalent widths can be obtained by multiplying H$\alpha$/R by
1262\,\AA\ (the equivalent width of the R filter).
We also measured the $F_{24}/F_8$ ratio at the peak of the 24\,$\mu$m emission
(which can be offset from both the exciting clusters and the peak of the
H$\alpha$ emission), within a resolution element (2.5 times the 24\,$\mu$m
FWHM, except for confused regions for which the aperture was reduced to
the FWHM in order to avoid neighboring sources), after convolving
the 8\,$\mu$m map to the 24\,$\mu$m angular resolution.

\begin{deluxetable}{lrrrrrr}
\tabletypesize{\footnotesize}
\tablecaption{List of selected clusters, named according to the associated
\ion{H}{2} region in the catalog of \citet{Deharveng}. A $\#$ symbol after
the name of an object indicates association with a supernova remnant in
the radio/X-ray catalogs of \citet{Pannuti} and \citet{Payne}, and a
$*$ symbol indicates association with a supernova remnant in the optical
catalog of \citet{Blair}.
\label{tab_clusters}
}
\tablehead{
name & RA~~~~~~~DEC~~ (J2000)                         & $D_1$~\tablenotemark{a} & $F$(H$\alpha$)            &
   $A$(H$\alpha$) & H$\alpha$/R & $F_{24} / F_8$ \\
~    & (h m s)~~(\degr\ \arcmin\ \arcsec)~~~~~~~~~~~~ & (arcsec)                & ($10^{-18}$\,W\,m$^{-2}$) &
   (mag)          & ~           & (peak)         \\
}
\startdata
D6            & $00~54~16.5~~-37~34~56$ & 18.0 & 139. & ...  & 0.44 & 3.32     \\
D7-8          & $00~54~17.1~~-37~35~19$ & 33.6 & 652. & ...  & 0.15 & 1.93     \\
D24~$*$       & $00~54~28.7~~-37~41~36$ & 24.0 & 549. & 0.38 & 0.38 & 1.30     \\
D39~$\#$      & $00~54~38.5~~-37~41~44$ & 12.0 & 172. & 0.82 & 0.87 & 1.76     \\
D40~$\#$      & $00~54~38.5~~-37~42~41$ & 14.4 & 173. & 0.15 & 0.28 & 0.94     \\
D45~$*$~$\#$  & $00~54~40.5~~-37~40~55$ & 10.8 & 363. & 0.81 & 0.41 & 0.93     \\
D53a~$*$~$\#$ & $00~54~42.9~~-37~43~12$ & 6.0  & 156. & 0.79 & 1.11 & 3.42     \\
D53b~$*$~$\#$ & $00~54~43.5~~-37~43~11$ & 8.4  & 365. & 0.69 & 0.80 & 2.64     \\
D53c~$*$~$\#$ & $00~54~42.9~~-37~43~00$ & 10.8 & 380. & 0.39 & 0.11 & $\sim 1$ \\
D56           & $00~54~44.3~~-37~40~27$ & 9.6  & 109. & 0.85 & 0.70 & 1.00     \\
D61~$\#$      & $00~54~45.4~~-37~38~47$ & 12.0 & 255. & 0.62 & 0.75 & 1.21     \\
D76a~$\#$     & $00~54~50.4~~-37~40~30$ & 8.4  & 123. & 0.88 & 0.69 & 0.97     \\
D76b          & $00~54~51.0~~-37~40~27$ & 7.2  & 93.  & 0.59 & 0.43 & 1.31     \\
D77~$\#$      & $00~54~50.4~~-37~38~24$ & 12.0 & 407. & 0.38 & 0.47 & 2.07     \\
D79~$\#$      & $00~54~51.2~~-37~38~26$ & 16.8 & 323. & 0.55 & 0.46 & 1.76     \\
D84~$\#$      & $00~54~51.8~~-37~39~39$ & 12.0 & 316. & 0.66 & 0.90 & 2.22     \\
D88-90        & $00~54~53.4~~-37~43~47$ & 33.6 & 435. & 0.51 & 0.32 & 1.36     \\
D115          & $00~55~02.6~~-37~38~28$ & 36.0 & 327. & 0.37 & 0.04 & 1.21     \\
D118~$\#$     & $00~55~04.1~~-37~42~51$ & 30.0 & 753. & 0.65 & 0.33 & 2.27     \\
D119~$\#$     & $00~55~03.9~~-37~43~20$ & 26.4 & 985. & 0.49 & 0.47 & 2.22     \\
D122          & $00~55~04.7~~-37~40~58$ & 14.4 & 149. & 1.06 & 0.63 & 1.06     \\
D127          & $00~55~07.6~~-37~41~51$ & 10.8 & 104. & 0.70 & 0.59 & 0.94     \\
D129~$*$      & $00~55~08.9~~-37~39~31$ & 15.6 & 109. & 0.58 & 0.51 & 0.99     \\
D137a~$\#$    & $00~55~13.0~~-37~41~41$ & 13.2 & 919. & 0.41 & 0.27 & 3.43     \\
D137b~$*$     & $00~55~12.4~~-37~41~26$ & 28.8 & 938. & 0.57 & 0.20 & 1.21     \\
D147          & $00~55~24.7~~-37~39~33$ & 26.4 & 148. & ...  & 0.10 & 1.04     \\
D159~$*$~$\#$ & $00~55~33.9~~-37~43~15$ & 24.0 & 196. & ...  & 0.31 & ...      \\
\enddata
\tablenotetext{a}{
This is the diameter encompassing the pixels selected
for the measurement of the cluster fluxes (see Section~\ref{photom} for details
on the photometric procedure).}
\end{deluxetable}

Table~\ref{tab_clusters} also indicates the association of the \ion{H}{2}
regions with supernova remnants in the catalogs of \citet{Pannuti},
\citet{Payne} and \citet{Blair}. We fail to see any direct connection
between the presence of a supernova remnant and the morphology of the
\ion{H}{2} region, as some of these are classified here as compact
(see in Section~\ref{extlaw} the discussion about a quantification of the
compactness of the \ion{H}{2} regions, and the significance of this parameter
in interpreting the results). Among these catalogued remnants, weak X-ray
sources are found close to D40, D45, and D53~; the radio sources are all
very weak, and some have flat radio spectral indices, consistent with
thermal emission. Such is the case for the regions D61, D77, D84.
The influence of these candidate young supernova remnants on the
surrounding interstellar medium is thus unclear, and the shell-like
structures visible in the H$\alpha$ emission may correspond to older events.

\subsection{Photometry}
\label{photom}

Special care was devoted to refine the photometric method so as to obtain
accurate spectral energy distributions for the young clusters.
In order to remove the local background from old populations, we devised
a method appropriate to the combination of UV and optical data.
We had to take into account the facts that the underlying emission can
vary significantly on the scale of the considered regions, that its
brightness relative to that of the clusters is variable from one band to
the other, increasing with wavelength and being very high in the R and
I bands, and that confusion makes any simple aperture always contaminated
by old stellar populations and foreground stars.
We define two diameters, constraining the source to be contained within
the smaller ($D_1$) and estimating the background within the larger ($D_2$).
Using the U band, where the background and contamination by old populations
are minimal, we then determine refined photometric ``apertures'' in the
following way. The average brightness $b$ and standard deviation $\sigma$
of the background in $D_2$ are computed iteratively in a narrowing
brightness interval ($\pm 2 \sigma_{\rm i}$ where $\sigma_{\rm i}$ is
the standard deviation at each step) until convergence is achieved. Then,
the source flux is summed over the pixels above $b + 3 \sigma$ within $D_1$
and the background averaged over the pixels below $b + 3 \sigma$ within $D_2$
(not excluding $D_1$).
Then, we use the exact same pixels to compute the source flux and
background in all the other optical bands, which have been previously
aligned to within 0.2\,arcsec, and control the results visually
to ensure that no visible foreground star contaminates the aperture,
and that $D_1$ and $D_2$ were appropriately chosen.
We find that this method produces significantly better results than
by measuring the background in a concentric annulus or another
aperture centered on a nearby empty patch. For clusters that appear
single in the optical bands, we obtain good fits with a single-population
model, which is not the case when we use the alternative methods above.

\begin{deluxetable}{lccccccc}
\tabletypesize{\footnotesize}
\tablecaption{
Broadband UV-optical photometric measurements, including the
emission line contributions to the U and R bands. The given error bars
are the formal measurements uncertainties only, to which we added
10\% for the flux calibration uncertainty.
\label{tab_photom}
}
\tablehead{
name & FUV       & NUV & U & B & V & R & I \\
~    & ($\mu$Jy) & ~   & ~ & ~ & ~ & ~ & ~ \\
}
\startdata
D6     & $892 (\pm 1$\%)  & $819 (\pm 1$\%)  & $835 (\pm 2$\%)  & $603 (\pm 2$\%)  & $540 (\pm 2$\%)  & $575 (\pm 2$\%)  & $391 (\pm 4$\%) \\
D7-8   & $2419 (\pm 1$\%) & $2052 (\pm 1$\%) & $1442 (\pm 2$\%) & $1090 (\pm 2$\%) & $834 (\pm 2$\%)  & $795 (\pm 2$\%)  & $611 (\pm 4$\%) \\
D24    & $1377 (\pm 1$\%) & $1315 (\pm 1$\%) & $1275 (\pm 1$\%) & $898 (\pm 1$\%)  & $730 (\pm 2$\%)  & $809 (\pm 2$\%)  & $482 (\pm 4$\%) \\
D39    & $287 (\pm 2$\%)  & $257 (\pm 2$\%)  & $392 (\pm 2$\%)  & $205 (\pm 3$\%)  & $166 (\pm 4$\%)  & $288 (\pm 5$\%)  & $110 (\pm 11$\%) \\
D40    & $510 (\pm 1$\%)  & $489 (\pm 1$\%)  & $439 (\pm 2$\%)  & $309 (\pm 3$\%)  & $234 (\pm 4$\%)  & $272 (\pm 4$\%)  & $133 (\pm 11$\%) \\
D45    & $526 (\pm 1$\%)  & $613 (\pm 1$\%)  & $862 (\pm 2$\%)  & $574 (\pm 2$\%)  & $472 (\pm 2$\%)  & $645 (\pm 3$\%)  & $326 (\pm 5$\%) \\
D53a   & $246 (\pm 1$\%)  & $196 (\pm 2$\%)  & $322 (\pm 2$\%)  & $153 (\pm 2$\%)  & $141 (\pm 3$\%)  & $258 (\pm 3$\%)  & $131 (\pm 6$\%) \\
D53b   & $733 (\pm 1$\%)  & $632 (\pm 1$\%)  & $913 (\pm 1$\%)  & $519 (\pm 1$\%)  & $452 (\pm 2$\%)  & $712 (\pm 2$\%)  & $376 (\pm 3$\%) \\
D53c   & $1387 (\pm 1$\%) & $1390 (\pm 1$\%) & $1182 (\pm 1$\%) & $971 (\pm 1$\%)  & $849 (\pm 1$\%)  & $933 (\pm 2$\%)  & $846 (\pm 3$\%) \\
D56    & $225 (\pm 2$\%)  & $200 (\pm 2$\%)  & $276 (\pm 2$\%)  & $165 (\pm 4$\%)  & $134 (\pm 6$\%)  & $213 (\pm 8$\%)  & $80 (\pm 18$\%) \\
D61    & $398 (\pm 1$\%)  & $356 (\pm 1$\%)  & $518 (\pm 2$\%)  & $301 (\pm 2$\%)  & $242 (\pm 3$\%)  & $403 (\pm 4$\%)  & $181 (\pm 9$\%) \\
D76a   & $310 (\pm 1$\%)  & $276 (\pm 2$\%)  & $338 (\pm 4$\%)  & $203 (\pm 5$\%)  & $148 (\pm 10$\%) & $246 (\pm 12$\%) & $145 (\pm 14$\%) \\
D76b   & $360 (\pm 1$\%)  & $301 (\pm 1$\%)  & $274 (\pm 3$\%)  & $189 (\pm 5$\%)  & $147 (\pm 8$\%)  & $214 (\pm 11$\%) & $132 (\pm 19$\%) \\
D77    & $1202 (\pm 1$\%) & $1082 (\pm 1$\%) & $906 (\pm 3$\%)  & $606 (\pm 4$\%)  & $503 (\pm 4$\%)  & $700 (\pm 5$\%)  & $419 (\pm 11$\%) \\
D79    & $929 (\pm 1$\%)  & $817 (\pm 1$\%)  & $908 (\pm 2$\%)  & $593 (\pm 2$\%)  & $482 (\pm 3$\%)  & $717 (\pm 3$\%)  & $398 (\pm 5$\%) \\
D84    & $411 (\pm 2$\%)  & $354 (\pm 2$\%)  & $621 (\pm 2$\%)  & $381 (\pm 2$\%)  & $341 (\pm 4$\%)  & $535 (\pm 5$\%)  & $275 (\pm 9$\%) \\
D88-90 & $1183 (\pm 1$\%) & $1252 (\pm 1$\%) & $1273 (\pm 1$\%) & $880 (\pm 2$\%)  & $700 (\pm 2$\%)  & $877 (\pm 3$\%)  & $491 (\pm 5$\%) \\
D115   & $1713 (\pm 1$\%) & $1766 (\pm 1$\%) & $1516 (\pm 1$\%) & $1529 (\pm 1$\%) & $1332 (\pm 2$\%) & $1329 (\pm 2$\%) & $1326 (\pm 3$\%) \\
D118   & $2594 (\pm 1$\%) & $2587 (\pm 1$\%) & $2294 (\pm 1$\%) & $1688 (\pm 2$\%) & $1372 (\pm 3$\%) & $1645 (\pm 3$\%) & $914 (\pm 5$\%) \\
D119   & $2887 (\pm 1$\%) & $2771 (\pm 1$\%) & $3007 (\pm 1$\%) & $1887 (\pm 1$\%) & $1401 (\pm 2$\%) & $1882 (\pm 2$\%) & $894 (\pm 6$\%) \\
D122   & $252 (\pm 2$\%)  & $239 (\pm 2$\%)  & $369 (\pm 3$\%)  & $232 (\pm 4$\%)  & $177 (\pm 6$\%)  & $292 (\pm 8$\%)  & $153 (\pm 15$\%) \\
D127   & $163 (\pm 2$\%)  & $139 (\pm 2$\%)  & $212 (\pm 3$\%)  & $137 (\pm 5$\%)  & $119 (\pm 7$\%)  & $188 (\pm 8$\%)  & $138 (\pm 8$\%) \\
D129   & $369 (\pm 2$\%)  & $309 (\pm 2$\%)  & $354 (\pm 2$\%)  & $210 (\pm 3$\%)  & $161 (\pm 3$\%)  & $245 (\pm 4$\%)  & $105 (\pm 10$\%) \\
D137a  & $4124 (\pm 1$\%) & $3784 (\pm 1$\%) & $2966 (\pm 1$\%) & $2323 (\pm 1$\%) & $1853 (\pm 1$\%) & $1900 (\pm 1$\%) & $1182 (\pm 2$\%) \\
D137b  & $3397 (\pm 1$\%) & $3516 (\pm 1$\%) & $3625 (\pm 1$\%) & $2913 (\pm 1$\%) & $2446 (\pm 1$\%) & $2744 (\pm 1$\%) & $2046 (\pm 2$\%) \\
D147   & $925 (\pm 1$\%)  & $835 (\pm 1$\%)  & $615 (\pm 1$\%)  & $560 (\pm 1$\%)  & $433 (\pm 2$\%)  & $410 (\pm 2$\%)  & $276 (\pm 7$\%) \\
D159   & $696 (\pm 1$\%)  & $664 (\pm 1$\%)  & $554 (\pm 2$\%)  & $354 (\pm 3$\%)  & $296 (\pm 5$\%)  & $406 (\pm 5$\%)  & $228 (\pm 11$\%) \\
\enddata
\end{deluxetable}

Since the point spread function of the {\small GALEX} images varies across
the field of view and is slightly elliptical in the FUV band, and because of
residual distortion between the FUV and NUV images, we had to refine the
photometric technique for the {\small GALEX} bands. In each small field
around a selected cluster, we use point sources to fit positional offsets
with respect to the optical images, and we determine the FWHM of
the cluster independently in the FUV and NUV bands. In a few
instances, where the selected cluster is partially confused by nearby
clusters, we fitted and removed the latter before performing aperture
photometry on the target. In a few cases when we have to truncate the
aperture to avoid confusion, we apply an aperture correction, computed
assuming that the source is perfectly gaussian.
The measured broadband fluxes are listed in Table~\ref{tab_photom}.

\subsection{Stellar population fits}
\label{text_fits}

We used the Starburst99 synthesis model \citep{Leitherer} to compute
spectra of instantaneously-formed stellar clusters, with the metallicity
$Z = 0.008$, well suited to NGC\,300, and a Salpeter initial mass function
between 0.1 and 120\,M$_{\odot}$.

The transmission curves of the filters used in the observations were applied
to the model spectra before doing any comparison between the models and
the measured fluxes. For the optical filters, we used the definitions of
the Johnson-Cousins system given by \citet{Fukugita} and the references
therein. For the ultraviolet and infrared filters, we used the actual
transmission curves of {\small GALEX}\footnote{available from the website:
{\it http://galexgi.gsfc.nasa.gov/tools/Resolution\_Response/index.html\,.}}
and Spitzer\footnote{available from:
{\it http://ssc.caltech.edu/irac/spectral\_response.html}  \\
and {\it http://ssc.caltech.edu/mips/spectral\_response.html\,.}}.

We fit the observed FUV-NUV-UBVRI flux densities using three alternate
extinction laws -- \citet{Cardelli} for the Milky Way,
with the ratio of total to selective extinction
$R_{\rm V} = A{\rm (V)} / E{\rm (B-V)} = 3.1$, appropriate for the diffuse
interstellar medium~;
\citet{Fitzpatrick} for
the LMC average and for the 30\,Doradus complex -- and the attenuation law
of \citet{Calzetti94}. The results should be interpreted while keeping
in mind that the first three extinction curves were derived toward the
line of sight of individual stars, and the last one is in principle
only valid for the integrated emission of starburst galaxies, while
we are applying them to
spatial extractions within
regions ranging between 60 and 300\,pc in diameter.
While \citet{Misselt99} have shown that the type of extinction law attributed
by \citet{Fitzpatrick} to 30\,Doradus is in fact mostly observed in a region
adjacent to 30\,Doradus, the LMC2 giant shell, we retain the terminology of
\citet{Fitzpatrick} for reference to the particular extinction curve
that he derived. Updated extinction curves for the LMC have been published
by \citet{Misselt99}. Given that we are interested in extracting qualitative
variations in the apparent local extinction law (see Section~\ref{extlaw}),
and that the discrete extinction curves used here are in fact part of a
continuum in dust properties \citep{Gordon03}, these curves are sufficient
for our purposes.

We fit the cluster spectral energy distributions in a simple dust screen
configuration, which is implicit in the four extinction curves considered
here, i.e. ignoring the effects of scattering and radiative transfer through
layers of dust mixed with the stars. These effects are extensively described
by e.g. \citet{Witt92} and \citet{Calzetti01}. Allowing the relative geometry
of stars and dust to be unconstrained would introduce insurmountable degeneracy
with the fitted parameters of the stellar populations. But there is another
motivation for assuming that the stars are not mixed with the dust. Our
photometric technique, in effect, is designed to extract the emission from
discrete stellar clusters and remove efficiently background or foreground
extended emission (Section~\ref{photom}). We argue that the stellar clusters
can thus be treated as a collection of point sources, each with its own dust
shell. This is especially true of the class of clusters that we call compact
(see Section~\ref{extlaw} for a quantitative definition). The smallest full
width at half maximum of the images represents a linear distance of 20\,pc,
to be compared with the full widths at half maximum of young clusters in
spiral galaxies as quantified by \citet{Larsen}, up to 10\,pc.
Even though the dust screen assumption does not reflect the likely geometry
of the sources, we will show in Section~\ref{extlaw} that it is a good
approximation for shell-type sources, for which the inclusion of radiative
transfer effects (in particular scattering effects) does not alter our
conclusions. Furthermore, the results concerning the variations of the
attenuation law (which includes radiative transfer effects, as opposed to
the pure extinction law) are independent of any assumptions about the source
geometry.
Note that scattering should be a negligible component
of the \citet{Calzetti94} law, because light scattered in the line of sight
compensates on average light scattered out of the line of sight for regions
of large size (several kiloparsec).

The flux calibration uncertainty is assumed to be 10\% for all the filters,
to which the formal measurement uncertainty is added. The fits were not constrained
to reproduce the 3-5\,$\mu$m data, but only not to overproduce them (within
the error bars), even when these bands are dominated by stellar emission.
We find in some cases that no single coeval population can reproduce
the photometric data, in which case we allow two populations of different
ages but same extinction. This occurs for the double cluster D53c
and for some complex fields associated with very large \ion{H}{2} regions
(D137b, D115). All the other regions were fitted with a single population.

In \ion{H}{2} regions, the H$\alpha$ and [\ion{O}{2}] lines at 6563\,\AA\
and 3727\,\AA, respectively, contribute a large fraction of the total flux
in the R and U bands (up to 50\% in R, and up to 40\% in U, respectively,
within the chosen apertures), except in very diffuse \ion{H}{2} regions.
The contribution of the H$\beta$ and [\ion{O}{3}] lines can be neglected
because they are observed through low-transmission parts of the B and V
filters, and are intrinsically weaker. We accounted for H$\alpha$ and
[\ion{O}{2}] fluxes by subtracting scaled versions of the H$\alpha$ map
from the R and U maps, using the [\ion{O}{2}]/H$\alpha$ ratios measured
by \citet{Webster} and \citet{d'Odorico} and correcting them for differential
extinction. Whenever a measurement of the [\ion{O}{2}]/H$\alpha$ ratio is
lacking, we adopt the average ratio of the \ion{H}{2} regions discussed
by \citet{Webster} and \citet{d'Odorico} ([\ion{O}{2}]/H$\alpha = 1.13$).
Because the [\ion{O}{2}] contribution estimated in this way is uncertain and
because of possible aperture mismatch, we affect a smaller weight to
the U-band data than to the other bands, recovering an updated
[\ion{O}{2}]/H$\alpha$ ratio from the SED fits (Table~\ref{tab_fits}).

The output parameters of the fits are the age(s) of the clusters, their
mass(es), and the monochromatic extinction at 1516\,\AA, the effective
wavelength of the FUV filter. The extinction and ages take discrete values,
varying in steps of 0.1\,mag and 1\,Myr, respectively. For each extinction,
each age of the young clusters, and optionally each age of older clusters,
the masses have to be solved from a set of 7 equations (for the 7 bands
from the FUV to I). In order not to bestow a larger weight on any particular
combination of bands, we solve separately each equation with the new equation
resulting from the sum of all the fluxes, and then we compute the average
of the 7 solutions, with a lower weight for the U band. The ionizing photon
flux predicted by the best fit is compared with the ionizing photon flux
estimated from the H$\alpha$ flux, corrected for extinction using the
H$\alpha$/H$\beta$ decrement, but is left unconstrained by the observations.
We usually find agreement to within 20\%, with a few exceptions
(Table~\ref{tab_fits}). To determine the confidence intervals of the
cluster parameters, the various fits are only constrained to reproduce
the ionizing photon flux from the best fit to within 30\% (we set a
constraint on the error, but not on the absolute value).

\begin{deluxetable}{lrrrrrlll}
\tabletypesize{\footnotesize}
\tablecaption{Results of the population synthesis fitting.
\label{tab_fits}
}
\tablehead{
name & $A$(1516\,\AA) & age~\tablenotemark{a} & mass                  &
   age~\tablenotemark{b} & mass~\tablenotemark{b} &
   [\ion{O}{2}]~\tablenotemark{c} &
   $N_{\rm Lyc}^{\rm model}$~\tablenotemark{d} &
   $F_{3-5\,\mu{\rm m}}^*$           \\
~    & (mag)        & (Myr)                   & ($10^3$\,M$_{\odot}$) &
   (2)                   & (2)                    &
   / H$\alpha$                    &
   $/ N_{\rm Lyc}^{\rm mes}$                   &
   $/ F_{3-5\,\mu{\rm m}}^{\rm tot}$ \\
}
\startdata
D6     & $0.2_{-0.1}^{+0.2}$ & $4_{-0}^{+1}$ & $10.2_{-0.5}^{+6.1}$ & ~   & ~    & 1.0/1.3       & 1.1-1.1-1.1-1.2 & $< 0.33$ \\
D7-8   & $0.0_{-0.0}^{+0.2}$ & 3             & $15.1_{-0.1}^{+2.1}$ & ~   & ~    & 1.1$^{*}$/... & 0.9-0.9-0.9-1.0 & $\sim 1$ \\
D24    & $0.7_{-0.2}^{+0.1}$ & 3             & $15.9_{-1.6}^{+0.9}$ & ~   & ~    & 1.2/1.1       & ...-...-1.1-1.3 & $< 0.64$ \\
D39    & $1.0 \pm 0.2$       & 3             & $4.6_{-0.6}^{+0.4}$  & ~   & ~    & 1.1$^{*}$/1.9 & 0.7-0.6-...-... & $< 0.26$ \\
D40    & $0.5 \pm 0.1$       & 3             & $5.0 \pm 0.3$        & ~   & ~    & 1.1$^{*}$/1.2 & ...-...-1.3-1.6 & $< 0.86$ \\
D45    & $1.5_{-0.1}^{+0.2}$ & 3             & $13.4_{-0.7}^{+1.3}$ & ~   & ~    & 1.5/1.8       & ...-...-1.0-... & $< 0.86$ \\
D53a   & $1.0 \pm 0.2$       & 1-2           & $3.9_{-0.3}^{+0.5}$  & ~   & ~    & 1.4/1.5       & 1.0-1.2-...-... & $< 0.24$ \\
D53b   & $1.1_{-0.2}^{+0.1}$ & 3             & $12.4_{-1.3}^{+0.8}$ & ~   & ~    & 1.4/1.4       & 0.9-...-...-... & $< 0.24$ \\
D53c   & $0.6 \pm 0.2$       & 3             & $13.0_{-2.2}^{+3.6}$ & 13  & 20   & 1.4/...       & ...-...-1.3-... & $\sim 1$ \\
D56    & $1.0 \pm 0.2$       & 3             & $3.6_{-0.4}^{+0.5}$  & ~   & ~    & 1.1$^{*}$/1.7 & 0.8-0.8-...-... & $< 0.17$ \\
D61    & $1.1_{-0.2}^{+0.1}$ & 3             & $6.9_{-0.8}^{+0.4}$  & ~   & ~    & 1.1$^{*}$/1.4 & 0.8-0.8-...-... & $< 0.41$ \\
D76a   & $0.8_{-0.1}^{+0.2}$ & 3             & $4.1_{-0.3}^{+0.6}$  & ~   & ~    & 1.6/1.7       & 0.9-0.8-...-... & $< 0.17$ \\
D76b   & $0.4 \pm 0.1$       & 3             & $3.2_{-0.2}^{+0.3}$  & ~   & ~    & 1.1$^{*}$/0.8 & 1.2-1.2-...-... & $< 0.22$ \\
D77    & $0.4 \pm 0.1$       & 3             & $11.0_{-1.0}^{+0.8}$ & ~   & ~    & 0.9/0.5       & 1.0-1.0-1.0-1.1 & $< 0.27$ \\
D79    & $0.8_{-0.1}^{+0.2}$ & 3             & $12.4_{-1.0}^{+1.6}$ & ~   & ~    & 1.2/1.1       & 1.2-1.3-...-... & $< 0.36$ \\
D84    & $1.5_{-0.1}^{+0.2}$ & 3             & $10.3_{-0.5}^{+1.3}$ & ~   & ~    & 1.1$^{*}$/1.2 & 0.9-...-...-... & $< 0.25$ \\
D88-90 & $0.9 \pm 0.1$       & 3             & $17.1 \pm 0.9$       & ~   & ~    & 1.1$^{*}$/1.3 & ...-...-1.3-1.7 & $< 0.26$ \\
D115   & $0.7_{-0.7}^{+0.1}$ & $5_{-2}^{+0}$ & $36_{-28}^{+7}$      & 9   & 16   & 1.1$^{*}$/... & ...-...-0.9-1.5 & $< 0.63$ \\
D118   & $0.7 \pm 0.1$       & 3             & $31 \pm 2$           & ~   & ~    & 1.0/0.8       & ...-...-1.2-1.4 & $< 0.51$ \\
D119   & $0.6_{-0.1}^{+0.2}$ & 3             & $31_{-2}^{+4}$       & ~   & ~    & 1.3/1.7       & ...-...-1.1-1.3 & $< 0.60$ \\
D122   & $1.4_{-0.1}^{+0.2}$ & 3             & $5.8_{-0.3}^{+0.7}$  & ~   & ~    & 1.1$^{*}$/2.1 & ...-0.8-...-... & $< 0.17$ \\
D127   & $1.4 \pm 0.1$       & 3             & $3.7 \pm 0.2$        & ~   & ~    & 1.1$^{*}$/1.2 & 1.0-...-...-... & $< 0.23$ \\
D129   & $0.6_{-0.2}^{+0.1}$ & 3             & $3.9_{-0.5}^{+0.3}$  & ~   & ~    & 1.1$^{*}$/1.5 & 1.1-...-...-... & $< 0.34$ \\
D137a  & $0.4 \pm 0.1$       & 3             & $37_{-1}^{+2}$       & ~   & ~    & 0.8/0.4       & ...-1.4-1.4-1.4 & $< 0.86$ \\
D137b  & $0.9_{-0.4}^{+0.3}$ & $3_{-2}^{+0}$ & $45_{-38}^{+13}$     & 23  & 62   & 1.1$^{*}$/1.1 & ...-...-1.4-2.0 & $< 0.69$ \\
D147   & $0.0_{-0.0}^{+0.1}$ & 5             & $11.9_{-0.1}^{+0.8}$ & ~   & ~    & 1.1$^{*}$/... & 0.9-0.9-0.9-0.9 & $< 0.80$ \\
D159   & $0.4 \pm 0.1$       & 3             & $6.4_{-0.4}^{+0.3}$  & ~   & ~    & 1.1$^{*}$/0.7 & ...-...-1.2-1.5 & $< 0.40$ \\
\enddata
\vspace*{-0.2cm}
\tablenotetext{a}{When no error bar on the age is given, it means that the
$1 \sigma$ interval is smaller than the time step of 1\,Myr used in the fits.
\vspace*{-0.2cm}}
\tablenotetext{b}{Since the age and mass of secondary stellar populations
are always ill-constrained, this table indicates only the best-fit values.
\vspace*{-0.2cm}}
\tablenotetext{c}{The first value is the input (from \citet{Webster} or \citet{d'Odorico},
or the average ratio of the regions discussed in these papers when followed by a star)~;
the second value is the output of the best fit (not given if the H$\alpha$/R
ratio is below 0.15, too low to derive [\ion{O}{2}]/H$\alpha$ from the fits). All the
sub-regions of D53 were assumed to have the same [\ion{O}{2}]/H$\alpha$ ratio.
\vspace*{-0.2cm}}
\tablenotetext{d}{The ratio of the ionizing photon flux predicted by the model to that
estimated from the observations is given for the best fit of each applicable extinction law,
in the following order: Milky Way, LMC, 30\,Doradus, and \citet{Calzetti94}.
Whenever the 30\,Doradus and \citet{Calzetti94} laws both provide acceptable fits,
since the \citet{Calzetti94} law predicts an ionizing photon flux usually in excess,
the cluster parameters and their confidence intervals are derived only for
the 30\,Doradus law (the \citet{Calzetti94} law produces larger masses and higher
extinctions).
\vspace*{-4cm} \\
~}
\vspace*{2cm}
\end{deluxetable}

We expect the error on the observed ionizing photon flux to be large,
because of uncertainties on the H$\alpha / {\rm H}\beta$ decrement, because
of uncertainties in the extent of the region directly under the influence
of the ionizing stars, and because an unknown fraction of the ionizing photons
may leak out of the \ion{H}{2} regions. The few cases where the agreement
between the model value and the observed value is poorer than 20\% may be due
to local errors in the subtraction of the continuum and stellar absorption
from the H$\beta$ line, or to increased porosity of the surrounding interstellar
medium. One of the regions for which the best-fit model overpredicts the measured
ionizing photon flux by a large percentage is D137b (by 40\% for the
30\,Doradus law, and 100\% for the \citet{Calzetti94} law). This region also
stands out as the most peculiar in terms of morphology of the hydrogen gas,
suggesting that previous generations of stars in the recent past have injected
lots of mechanical energy in the surrounding medium, carving a tunnel-shaped
void. It can thus be hypothesized that a sizeable fraction of the ionizing
photons are leaking out of the visible \ion{H}{2} region. According to
studies of the diffuse ionized gas in spiral galaxies, up to 50\% of the
ionizing photons may escape \ion{H}{2} regions \citep{Ferguson}. Alternatively,
the inclusion of a second population increases the degeneracy of the fits,
and the fitted stellar extinction could be too high.

We assess the validity of the different extinction laws by first comparing
the $\chi^2$ values of the best fits. But since the absolute uncertainty
was fixed arbitrarily to 10\%, not taking into account the smaller relative
uncertainty between the two {\small GALEX} bands, we consider that a $\chi^2$
value close to unity does not necessarily indicate that the fit is acceptable.
We also check its ability to reproduce colors, in particular the FUV-NUV
color, which is the most useful discriminant between the different extinction
laws. The ratio of the extinction in these two bands varies from $\sim 1.00$
for the Milky Way and $\sim 1.10$ for the LMC to $\sim 1.27$ for 30\,Doradus
and $\sim 1.20$ for the \citet{Calzetti94} law.

Whenever some significant interstellar component at 3-5\,$\mu$m was
indicated by the population synthesis model, we confirmed this
finding by examining the morphology of the dust and stellar clusters
(i.e. by checking that the structure of the 3-5\,$\mu$m emission is
much more similar to that at 8\,$\mu$m than to that of the stars).

\section{Results and discussion}
\label{discussion}

\subsection{Ages}

One characteristic of the fits assuming instantaneous star formation
is that they nearly always predict a cluster age of 3\,Myr, with the
notable exception of D53a, which is fitted with a stellar population
of 1 to 2\,Myr, and is one of the regions with the highest $F_{24}/F_8$
flux ratios. This uniformity of age is likely an artifact of adopting
too simplified a representation of the clusters (instantaneous star
formation model with coarse time steps of 1\,Myr), whereas their formation
may go on for several Myr instead of being coeval. That the clusters
are all young and observed before the bulk of supernova explosions can be
inferred from the fact that they produce a significant amount of ionizing
radiation, since most were selected to be associated with bright
\ion{H}{2} regions. The case of D53a is interesting, as it confirms that
complexes containing younger stars tend to have higher ratios of very small
grain emission to aromatic band emission \citep{Sauvage, Roussel01b, Dale04}.

\subsection{Masses}

When considering such small stellar populations as those extracted here,
formed of one or a few young clusters, statistical fluctuations of the
mass spectrum of massive stars may become severe enough to affect the
results of the fits. For the adopted initial mass function, the ratio
${\rm N} ({\rm M} > {\rm M}_{\rm hot}) / {\rm M}_{\rm tot}$, where the
numerator is the number of stars more massive than the threshold M$_{\rm hot}$
and the denominator is the total cluster mass as given in Table~\ref{tab_fits},
is equal to $0.1264 {\rm M}_{\rm hot}^{-1.35} - 0.0002$, with masses
in units of solar mass. The whole range of cluster masses in
Table~\ref{tab_fits} thus corresponds to between 9 and 126 O stars
with ${\rm M}_{\rm hot} = 16\,{\rm M}_{\sun}$ \citep{Vacca}. Adopting as
a working definition of ionizing stars ${\rm M}_{\rm hot} = 10\,{\rm M}_{\sun}$
\citep{Kennicutt84}, the clusters masses also correspond to between
17 and 245 ionizing stars. The sampling of the initial mass function is
thus a concern for the smallest clusters analysed in this paper, in which
a single ionizing star may account for up to 5\% of the far-ultraviolet
emission. Note that a few regions, for which we could not obtain any
satisfactory fits, were discarded from our sample~; one of the possible
reasons for this is small-number statistics of ionizing stars in low-mass
stellar clusters.

\subsection{Stellar and nebular extinctions}

The fitted stellar extinction at the effective wavelength of the FUV
filter varies between nearly zero to $\sim 1.5$\,mag. Its correspondence
with the nebular extinction is shown in Figure~\ref{fig:a_ha_a1516}.
The lack of a tight correlation between the two can be understood in
terms of spatial offsets between \ion{H}{2} regions and clusters,
and highly variable geometric configuration from one region to another.
Within this sample, the A(H$\alpha$)/A$_{\rm R}$ ratio ranges from
1 to more than 3.5\,. The deviation from the value usually adopted for
actively star-forming galaxies \citep{Calzetti94} can thus be greater
than $\pm 50$\% for individual regions.

There is a good qualitative correspondance between the extinction map at
a resolution of 38\arcsec\ and the extinctions of individual regions
derived from the population synthesis fits, considering that significant
deviations are expected from the fact that extinction is highly variable
on small scales. It should also be noted that while the population
synthesis fits give the extinction of the young stellar clusters only,
the extinction in Figure~\ref{fig:a_fuv}, by construction, includes all
the stellar populations at any given location.
Since intermediate-age and old populations are, on average, much less
affected by extinction at any wavelength, because decoupled over time
from their parent gas clouds, the extinction values derived from the
global energy budget are expected to be much lower than those of the
ionizing clusters~; this is indeed observed.

\subsection{Extinction law and geometry of the interstellar medium}
\label{extlaw}

We find clear differences in which extinction laws are applicable,
as a function of the geometry of the \ion{H}{2} regions with respect to
the exciting clusters. The most compact regions follow either the
Milky Way or the LMC law, whereas
the clusters associated with
\ion{H}{2} regions forming shells
and diffuse structures follow either the 30\,Doradus or the \citet{Calzetti94}
law. Usually, the degeneracy between the latter two cannot be lifted,
but models using the \citet{Calzetti94} attenuation law systematically
overpredict the ionizing photon flux by much larger amounts than the
30\,Doradus law. We therefore consider the 30\,Doradus law to yield
a more reliable representation of the data, even though the stellar SED
fits are formally of comparable quality using either of these two laws.

To qualitatively ascertain the consequences of neglecting radiative transfer
effects, we examine the predictions of the {\small DIRTY} model
\citep{Witt00, Gordon01, Misselt01} with empirically-derived dust properties,
in the configuration of a dust shell surrounding a stellar cluster, appropriate
for our sample (Section~\ref{text_fits}). The dust properties of the LMC2 giant
shell (called 30\,Doradus here), not included in the original models, were
derived by \citet{Gordon03}\footnote{The data tables were retrieved from the
{\small DIRTY} website: {\it http://dirty.as.arizona.edu/}\,.}. The results
obtained from the pure extinction laws for the Milky Way, the LMC2 region
and the SMC are compared with those obtained from the corresponding attenuation
laws for two types of local dust distribution: homogeneous and clumpy, the
latter using the parameters defined by \citet{Witt00}.
While other geometric configurations are available within {\small DIRTY},
better applicable to the extended emission of large regions within galaxies,
it is beyond the scope of this observational study to perform comprehensive
modelling. We stress that the basic result of large variations in the
attenuation law, whether they correspond to genuine variations in the
extinction law or not, is independent of any assumptions about the relative
geometry of stars and dust.

Figure~\ref{fig:att_laws} shows the extinction and attenuation curves
in the model shell configuration,
for a particular optical depth, which corresponds to $A_{1516} = 0.7$
using the Cardelli extinction law, within the range of derived extinction
values for our clusters. The inclusion of scattering produces emerging
spectra that are bluer, with significantly higher FUV/NUV ratios, as more
light is scattered in the line of sight than out. It is thus conceivable
that the clusters which were well fitted with the Cardelli extinction law
could be equally well fitted with a LMC or SMC-based attenuation law.
However, even at the low optical depth considered in Figure~\ref{fig:att_laws},
the attenuation curves applicable to the clumpy case are extremely grey,
and thus would not provide any good fit to the observed spectral energy
distributions, since they would predict too much intrinsic emission in the
optical bands.

We now examine the aptitude of each attenuation law to reproduce the observed
FUV/NUV flux density ratios. Figure~\ref{fig:fuv_nuv_predictions} shows the
predictions from the {\small DIRTY} model. The intrinsic FUV/NUV ratio of a
3\,Myr-old cluster is 1.18, and it varies between 1.13 and 1.31 for ages
between 1 and 5\,Myr. This uncertainty is reflected in the error bars in
Figure~\ref{fig:fuv_nuv_predictions}. \\
Cluters well fitted with the Cardelli or LMC extinction law: \\
The data points overlap only with the Milky Way extinction law and the
LMC2 extinction and attenuation laws. However, the attenuation law
using the LMC2 dust properties, in the homogeneous case, is not much
different from the pure extinction law, i.e. the effects of scattering are
almost negligible. It is in principle possible to reproduce the observed
FUV/NUV ratios with the LMC2 or 30\,Doradus laws, but only if the clusters
are younger (1 or 2\,Myr).
This seems unlikely, since the predicted ionizing photon fluxes would
then be in strong excess with respect to the observational estimates
(which are more robust for the compact H{\small II} regions associated
with these clusters than for the diffuse H{\small II} regions).
We note that the
effects of scattering are strongest for the Milky Way dust properties,
and that none of the clusters would be well represented in the dust
shell configuration with the Milky Way attenuation laws (i.e. including
scattering), either in the homogeneous or in the clumpy case. \\
Cluters well fitted with the 30\,Doradus extinction law: \\
The data points lie in between the LMC2 (30\,Doradus) and the SMC laws.
Again, the SMC-dust attenuation law, in the homogeneous case, is very
close to the pure extinction law, and the effects of scattering are
also very small in this configuration. But the extinction laws in
the line of sight of these clusters may span the full range between
the 30\,Doradus and the SMC laws. \\
We conclude that the correct treatment of radiative transfer effects
would not drastically alter our qualitative conclusions in what follows,
unless stars had extended spatial distributions and were significantly
mixed with the dust, which we argue is very unlikely for the stellar
clusters that we classify as compact.

To quantify the compactness of the regions in a convenient way, we use
the H$\alpha$/R flux ratio, measured in the area defining
the stellar aperture (i.e. if the \ion{H}{2} region extends far away
from the clusters, only a fraction of the H$\alpha$ emission is considered
in this ratio). While this is proportional to the H$\alpha$ equivalent
width, we argue below that it is also a valid measure of the compactness
of the \ion{H}{2} region. Although it is potentially distorted by
differential extinction between the nebular phase and the stars,
this effect should not have a large impact. Using the results from
Figure~\ref{fig:a_ha_a1516}, we derive that for the highest extinctions,
observed values of H$\alpha$/R close to 1 at one extreme or close to 0.2
at the other extreme may be shifted by up to 0.9 upward or 0.2 upward,
respectively, when corrected for extinction. Since the regions with the
highest extinctions are also those with the highest H$\alpha$/R values,
correcting rigorously for extinction would preserve the trend discussed
below. Figure~\ref{fig:ew_compactness} confirms that the H$\alpha$/R
ratio provides a reliable quantification of the degree of compactness
of the entire \ion{H}{2} region. We prefer to adopt the H$\alpha$/R
indicator, because it is a relatively unbiased measurement: unlike the
ordinate of Figure~\ref{fig:ew_compactness}, it does not imply any
assumption about the total extent of the \ion{H}{2} region, which is
difficult to define owing to the presence of neighboring \ion{H}{2}
regions and the diffuse and filamentary aspect of the H$\alpha$ emission.

Figure~\ref{fig:extlaw_ew} shows the dependence of the extinction
law on the H$\alpha$/R ratio. Whenever a region sits in between
two extinction laws, it means that its SED is formally as well represented
with either law as with the other. Unconstrained regions are those whose
SED is well fitted by at least three of the four extinction laws (usually
because of very small extinction). To indicate cases when the measured ionizing
flux is overestimated by a large amount by the \citet{Calzetti94} attenuation
law, the points are plotted closer to the 30\,Doradus law locus.
The main discriminant between the various extinction laws is the
A(FUV)/A(NUV) ratio (Sect.~\ref{text_fits}). The increase in
A(FUV)/A(NUV), going from the \citet{Cardelli} law to the LMC law to
the 30\,Doradus law, is due to both the gradual disappearance of the so-called
2175\,\AA\ bump, which is covered by the NUV band, and the steepening of
the FUV rise. Although the \citet{Calzetti94} law is devoid of the 2175\,\AA\
bump, the corresponding A(FUV)/A(NUV) ratio is slightly smaller than that
of the 30\,Doradus law, because the \citet{Calzetti94} law has a flatter
wavelength dependence.

That the extinction law should vary within a galaxy is not a
complete surprise given that some significant variations have been
observed within the Milky Way and the Magellanic Clouds.
In the SMC, \citet{Lequeux} discussed the line of sight to a star
with a Milky Way-type extinction law. In the LMC,
\citet{Misselt99} discussed a line of sight
associated with a giant shell, with a much weaker 2175\,\AA\ bump than
its immediate surroundings.
In the Milky Way, LMC-type extinction laws have be found toward
low-density lines of sight \citep{Clayton}, and a SMC-type extinction
law was derived toward a region suggested to have been profoundly
affected by supernova shocks \citep{Valencic}. Although the disappearance
of the 2175\,\AA\ bump, together with the steepening of the far-ultraviolet
rise, are usually ascribed to a decrease in metallicity, which could
explain the differences between the average extinction laws of the
Milky Way and Magellanic Clouds, \citet{Mas-Hesse} found no such dependence
of the extinction curve on metallicity, for a sample of dwarf star-forming
galaxies. As pointed out by \citet{Misselt99}, an apparent dependence on
metallicity may be only an indirect effect, because the dust grain
properties and geometry of the interstellar medium are expected to be
on average related to the metallicity.

It has been suggested that the intensity of star formation in the local
environment could be the main driver of the extinction law shape,
causing the destruction or processing of the dust grains responsible
for the 2175\,\AA\ feature, via radiation hardness and supernova shocks
\citep{FitzpatrickMassa, Gordon97, Gordon98, Mas-Hesse}. This scenario
is well in line with the attribution of part of the 2175\,\AA\ feature
to aromatic hydrocarbons \citep{Joblin, Duley}, as this dust species
is known to be less resilient than very small grains and to be destroyed
by intense radiation fields. Supporting this identification, \citet{Vermeij}
have reported variations of the mid-infrared band ratios (caused by
changes in molecular structure, or degree of ionization and hydrogenation)
that are correlated with the strength of the 2175\,\AA\ bump.
It has been proposed that aromatic hydrocarbons are also responsible
for part of the far-ultraviolet rise \citep{Puget}, and very small
carbonaceous grains (of three-dimensional structure) are sometimes
assumed to be the main contributor to the 2175\,\AA\ bump \citep{Draine,
Hecht, Sakata}.

The present work, however, supports the idea that geometry effects are
much more important than the distinction between quiescent and actively
star-forming regions. According to the SED fits, all the exciting clusters
have very similar ages. Therefore, rather than being a pure indicator of the
evolution of the clusters, as the equivalent widths of the H recombination
lines are usually assumed to be, the H$\alpha$/R ratio can be interpreted
here as mainly governed by the geometric evolution of the whole region,
bearing traces of previous generations of stars in the shaping of the
interstellar gas. Indeed, the giant shells of hydrogen that are observed
in many locations across the disk of NGC\,300, where they are illuminated
by ionizing clusters, may have been carved by supernova explosions, in which
case they signal past star formation at the same location, more than 5\,Myr
ago. The difference in geometry between the regions D53a-b and D53c, which
are part of the same complex and whose exciting clusters have approximately
the same age, is particularly significant. High $F_{24}/F_8$ ratios,
indicative of intense radiation fields, can also be found both in compact
and diffuse star formation complexes. The most obvious difference between
clusters associated with a Milky Way-LMC extinction law and those associated
with a 30\,Doradus-Calzetti law is in the distribution of the material
around them. For the latter, both the atomic hydrogen and the
mid-infrared-emitting dust (aromatic hydrocarbons and very small grains)
have been removed far away, presumably by stellar winds and supernova shocks
from previous generations of stars at the same location. This does not
preclude important differences as well in the processing of dust grains
by supernova shocks, but if the modified dust is dissipated together
with the gas, it is only weakly participating in the extinction of the
clusters.

We can speculate that the effects of a varying extinction law within
and among galaxies may resemble the effects of a varying initial mass
function \citep{Rosa}, since for a given optical depth, the emerging
ultraviolet continuum is much flatter in the case of a 30\,Doradus-Calzetti
law than for a Milky Way law. An upper mass cutoff of the initial mass
function is not supported in NGC\,300, since Wolf-Rayet features
have been observed throughout its disk \citep{Schild}.

Finally, we can expect the behavior of more massive galaxies and of
starbursts, where \ion{H}{2} regions may generally be more centrally
distributed and more compact than in NGC\,300, to be significantly different.

\subsection{Interstellar emission in the near-infrared}

In some regions, we detect a clear excess at 3-5\,$\mu$m above the stellar
model. The improvement with respect to the previous treatment in
\citet{Helou04} (where it was assumed that the 3.5\,$\mu$m emission
covered by the IRAC1 band is entirely stellar) allows us to derive
more accurately the amount of interstellar emission at these wavelengths.
However, our method restricts the detectable interstellar component at
3-5\,$\mu$m to the exact location of the stellar emission~; if the hot
dust in this wavelength range is significantly offset from the exciting
source, it will not be detected. The 3.5\,$\mu$m/4.5\,$\mu$m color of young
clusters is very similar to that of dust (close to 1), making it very
difficult to separate the two components if extensive information at
other wavelengths does not exist.
Complexes where the interstellar component at 3-5\,$\mu$m is high with
respect to the stellar fluxes (within the stellar aperture) are among
the most compact (Fig.~\ref{fig:extlaw_ism} and \ref{fig:extlaw_ew}).

To explore the link between the extinction law and the dust emission,
we adopt measurements of the 3-5\,$\mu$m interstellar emission, as
the favorable angular resolution in these bands allows us to select
the dust emission within the stellar ``apertures'', directly in the
line of sight of the clusters. Figure~\ref{fig:extlaw_ism} confirms
that switching from the H$\alpha$ equivalent width to the column density
of near-infrared-emitting dust in the line of sight of the clusters
gives the same trend as seen in Figure~\ref{fig:extlaw_ew}: the shape
of the extinction law seems to depend importantly on the amount of
interstellar material which re-emits the absorbed energy in the
near- to mid-infrared. We cannot explore related variations in the
mid-infrared to far-infrared colors, because the angular resolution
at far-infrared wavelengths is too coarse.

Even though this 3-5\,$\mu$m dust component may not be physically
related to the other mid-infrared dust species, the aromatic band carriers
and very small grains, Figure~\ref{fig:f8_f3} shows that it scales
reasonably well with the 8\,$\mu$m aromatic band emission. These
measurements suffer from the degraded angular resolution at 8\,$\mu$m
combined with the narrow stellar ``apertures'' used to fit and extract the
interstellar component from the 3.6\,$\mu$m band, but are nevertheless
in line with the $F_{3.6}/F_8$ flux ratios derived by \citet{Lu}
for interstellar emission in star-forming galaxies, as well as for
the general interstellar medium of the Milky Way.

\subsection{Porosity of the interstellar medium and excitation of aromatic band carriers}

Examination of the morphology of the \ion{H}{2} complexes in H$\alpha$
provides evidence that the interstellar medium is highly porous in some regions,
thus that the 8\,$\mu$m-emitting dust can be heated by photons travelling
as far as several hundred parsecs (since the NUV photons travel farther
away than the ionizing photons, which are able to excite H$\alpha$ emission
a few hundred parsec away from the clusters).
Diffuse emission also exists at 24\,$\mu$m, but is harder to detect,
because it is intrinsically fainter than at 8\,$\mu$m and observed at
much lower angular resolution, increasing the confusion by nearby
bright sources, and the 8\,$\mu$m observations are deeper. Although the
8\,$\mu$m/H$\alpha$ flux ratio varies dramatically from bright \ion{H}{2}
regions to more diffuse regions \citep{Helou04}, and the 8\,$\mu$m emission
has a more filamentary aspect than the H$\alpha$ and 24\,$\mu$m emission,
virtually every 8\,$\mu$m peak is associated with a close-by H$\alpha$ peak
(see Figure~\ref{fig:images}).

The claim of \citet{Haas02} that aromatic band emission is more closely
correlated with very cold dust than with hot dust within galaxies
does not withstand examination in NGC\,300. Figure~\ref{fig:pah_bg} shows
the relation between the aromatic band emission at 8\,$\mu$m and the
emission in the far-infrared, at 70 and 160\,$\mu$m. As a comparison,
the relation between 24\,$\mu$m and the far-infrared emission is also
included. The 160\,$\mu$m emission can be considered a tracer of the
cold dust, and the 70\,$\mu$m emission arises from dust that is on average
hotter,
with a significant contribution from impulsively-heated very small grains
\citep{Desert}.
Figure~\ref{fig:pah_bg} indicates clearly that aromatic bands
are not associated with cold dust, and are not excited by the same
radiation as the big grains. In the 8\,$\mu$m-70\,$\mu$m diagram,
at high surface brightnesses, a branch with a slightly lower slope than
in the low-brightness part is visible. This branch is slightly more prominent
in the 8\,$\mu$m-24\,$\mu$m diagram~; it is linked with the disappearance
of the aromatic bands from inside the \ion{H}{2} regions. Nevertheless,
the 8\,$\mu$m-70\,$\mu$m and 8\,$\mu$m-24\,$\mu$m relations are much
closer to linearity than the 8\,$\mu$m-160\,$\mu$m relation.
The linearity between 8\,$\mu$m and 850\,$\mu$m surface brightnesses
reported by \citet{Haas02}, obtained at low angular resolution in
galaxies of moderate angular size, and more importantly at mediocre
sensitivity at 850\,$\mu$m, may simply be a consequence of extracting
only the brightest and most active regions, which generally have
dust temperature distributions shifted toward higher values, and
discarding more diffuse regions, with cooler dust temperatures.
Their result thus does not imply that the aromatic bands and the cold
dust component are physically related, nor heated by the same radiation.

\subsection{Embedded and bare sources}
\label{extremes}

Within the inner disk, a few non-stellar sources bright at 24\,$\mu$m
have weak optical and no ultraviolet counterparts (Table~\ref{tab_embedded}).
Two of these are probably background distant galaxies, judging from
their morphology in optical high-resolution images. The five other
sources coincide with very faint H$\alpha$ emission, have no other
optical counterpart, and may be ultracompact \ion{H}{2} regions.
We consider it likely at least for the object catalogued by
\citet{Deharveng} as the faint and small \ion{H}{2} region D125.
If the H$\alpha$/H$\beta$ decrement of D125 can be trusted, it indicates
a lower limit of the extinction in the H$\alpha$ line of 2.8\,mag, much higher
than for the \ion{H}{2} regions associated with the clusters examined
here (see Table~\ref{tab_clusters}). Its mid-infrared counterpart has
the highest $F_{24}/F_8$ ratio within the galaxy ($\sim 4.5$) and is
still visible at 70\,$\mu$m despite the degraded angular resolution.

\begin{deluxetable}{ll}
\tablecaption{24\,$\mu$m sources with weak optical and no ultraviolet counterpart.
\label{tab_embedded}
}
\tablehead{
RA~~~~~~~~~~~DEC~~ (J2000)                         & identification \\
(h m s)~~~~~~(\degr\ \arcmin\ \arcsec)~~~~~~~~~~~~ & ~              \\
}
\startdata
00~54~41.7~~-37~40~22 & ultracompact \ion{H}{2} region~? \\
00~54~50.4~~-37~38~51 & background galaxy~? \\
00~54~50.7~~-37~41~55 & ultracompact \ion{H}{2} region~? \\
00~54~53.3~~-37~43~13 & background galaxy~? \\
00~54~59.8~~-37~39~27 & ultracompact \ion{H}{2} region~? \\
00~55~01.7~~-37~39~32 & ultracompact \ion{H}{2} region~? \\
00~55~05.9~~-37~42~14 & D125 -- ultracompact \ion{H}{2} region~? \\
\enddata
\vspace*{5cm}
\end{deluxetable}

Conversely, some far-ultraviolet and H$\alpha$ sources are not
associated with any detectable mid-infrared dust emission.
These regions are of low brightness and diffuse in the H$\alpha$
emission, suggesting that they are aged \ion{H}{2} regions,
excited by slightly evolved clusters (see the previous paragraph).
Both types of sources contribute negligibly to the emission from the
optically-bright disk, but the latter (bare sources) become predominant
in the outer disk.

\section{Summary and conclusion}
\label{summary}

The large-scale UV/IR ratio gradient observed in NGC\,300 is probably
the consequence of a combination of factors, all causing an increase
in the mean free path of ultraviolet photons. In this paper, we have
focussed on geometric factors, and shown that they cannot be neglected.

Throughout the disk of NGC\,300, mid-infrared peaks are commonly offset
from both the ionized gas peaks and stellar clusters, except for compact
regions \citep{Helou04}. This is because we are able to partially resolve
\ion{H}{2} regions from photodissociation regions. However, in the outer
disk, even the far-infrared emission sometimes appears significantly
displaced from the location of the ultraviolet sources, at a resolution of
38\arcsec\ (FWHM) or about 400\,pc. In the outer disk, dust may be removed
away from star-forming sites more efficiently than in the inner disk,
either because the volume density of the interstellar medium is decreased
or because the exciting clusters are older on average and have injected
more non-thermal energy into the interstellar medium.

There are several examples of very porous star-forming sites at distances
of 3-4\arcmin\ ($\sim 2$\,kpc) from the center, from which both the
hydrogen gas and the dust have been removed (such as D118 and D137)~;
as they belong to the optically-bright part of the disk, where presumably
the molecular gas density remains high, i.e. above the threshold for
gravitational instabilities discussed by \citet{Kennicutt89}, these individual
\ion{H}{2} complexes are unable to affect significantly the global
UV/IR radial profile. However, such regions may become dominant in the
outer disk.

Most of the H$\alpha$ emission in the outer disk is in the form
of diffuse, very large shells, which supports either one of the factors
invoked above for dust removal: either the gas is very rapidly pushed
away, during the lifetime of ionizing stars, because the interstellar
medium is more tenuous, or the exciting stellar populations are older
on average, so that they had more time to blow the gas away by their winds
and supernova explosions.
Good supporting examples of the hypothesis of aged stellar populations
are the clusters D115 and D147, which are bright in the FUV, but seem
to contain slightly evolved stellar populations (of the order of 5\,Myr,
as opposed to 3\,Myr for most clusters). Since it is not possible to
measure and fit the SEDs of a reasonably large number of clusters
associated with giant shells in the outer disk, however, we cannot
constrain the importance of this scenario relatively to the others
(but see also Figure~\ref{fig:fuv_nuv}).

From population synthesis fitting of individual young stellar clusters
or small groups of clusters, we find that the stellar attenuation
law varies significantly on the explored spatial scales of $\sim 100$\,pc,
and we argue that this empirical result likely corresponds to real variations
in the extinction law as well.
Just as the UV/IR ratio does not depend only on age and metallicity,
there is ample evidence that the apparent correlation between the shape
of the extinction law and the metallicity or the mode of star formation
(active versus quiescent) is only a secondary effect of a more direct
cause. We argue that the strength of the 2175\,\AA\ bump and the steepness
of the far-ultraviolet rise, whatever the responsible dust species,
are directly related to the degree of compactness of the star-forming
regions, and to the column density of interstellar material in front
of the clusters. While most of the young stellar clusters considered
here have similar ages and have not produced supernova explosions yet,
the only obvious difference between them is that some appear to have
formed in regions deeply affected by previous generations of stars.
These are associated with diffuse, shell-like \ion{H}{2} regions of
large dimensions. If we divide star-forming complexes in two categories,
based on the degree of compactness of the H$\alpha$ emission, we find
that the most compact regions follow either the Milky Way or the
average LMC extinction law, whereas the most diffuse regions follow
an extinction law with a decreased 2175\,\AA\ bump and/or a steeper
far-ultraviolet rise. While we cannot ascribe these features to a
particular dust species, we show, using the near-infrared dust component
extracted from our SED fitting procedure, thanks to the favorable
angular resolution, that the distinction between the two types of
morphologies and the two types of extinction laws is related to the
column density of near-infrared emitting dust lying in the line of
sight of the clusters.

We can hypothesize that the main difference in dust nature causing the
change of extinction law between the compact and diffuse regions is
related to a changing proportion of dust grains very close to the \ion{H}{2}
region, which have been processed by hard ultraviolet radiation from
the young stars, with respect to grains lying further away and which
have not been processed in the same way. This photo-processing is likely
occurring very fast, so that the observed dust properties of compact
\ion{H}{2} regions should be systematically different from the dust properties
of diffuse regions where dust lies further away from the exciting stars.
According to \citet{Boulanger}, who discussed the ubiquitous and large
variations in mid-infrared to far-infrared colors within and among nearby
molecular clouds, such variations cannot be explained only by changes in
the radiation field, but require actual variations in the abundance of
small grains with respect to big grains. They proposed that these
abundance variations are caused by photo-processing, on time scales
comparable to those of the mixing of gas by turbulent motions in clouds.
\citet{Habart} and \citet{Abergel} also discussed abundance variations
of small grains and aromatic hydrocarbons in resolved photodissociation
regions and molecular clouds.
This speculation would be consistent with the fact that the dust emitting
in the 3-5\,$\mu$m range is likely excited inside or very close to
\ion{H}{2} regions \citep{Helou04}, and that we have seen here that
its excitation has a close correspondence to the shape of the extinction law.

\appendix

\section{Notes on individual regions (including the clusters for which
the fits are shown in Figure~\ref{fig:fits})}
\label{annexe}

{\bf D53a:}
Compact, associated with very hot dust at 24\,$\mu$m and 3-5\,$\mu$m.
Since the emission is confused by nearby clusters in the {\small GALEX}
bands, we had to subtract a gaussian representation of these contaminating
clusters, and apply a large aperture correction to the FUV and NUV
fluxes, because of the small size of the aperture. \\

{\bf D53b:}
Compact, associated with hot dust at 24\,$\mu$m and 3-5\,$\mu$m.
Two distinct clusters can be distinguished, but are well fitted
with a single stellar population. \\

{\bf D53c:}
Two clusters, one with a very red SED. The morphology in the various
bands clearly indicates that interstellar emission is negligible
at 3-5\,$\mu$m. The fits included a second population but were not
constrained to reproduce the 3-5\,$\mu$m fluxes. The modelling confirms,
however, the stellar nature of the 3-5\,$\mu$m emission.
The ionized gas is diffuse and forms an extended fragmentary shell
of size $\sim 200$\,pc. \\

{\bf D118:}
Complex region with several clusters, which can be fitted by a single
population. The ionized gas and the dust are distributed in both
compact regions and a giant shell. \\

{\bf D119:} Same remarks as for D118. \\

{\bf D137b:}
This complex region stands out as very peculiar: the ionized gas and
dust seem to form two walls on each side of the brightest cluster,
separated by $\sim 80$\,pc in projected distance. Star formation has
thus had dramatic effects on the shaping of the surrounding interstellar
medium. The H$\alpha$ and dust peaks are coincident with weak UV sources,
presumably because of extinction effects, while the cluster dominating
the UV emission is located in a zone nearly void of interstellar emission.
Since the complex contains redder stellar sources besides these clusters,
we allow for a second population in the fits. This increases dramatically
the degeneracy of the younger population parameters. \\

{\bf D84:}
Compact region, associated with hot dust at 24\,$\mu$m and 3-5\,$\mu$m. \\

\clearpage

\acknowledgements
We thank Sang Chul Kim and Charles Hoopes for graciously providing their
UBVI and H$\alpha$ images, respectively, of NGC\,300.
We are most grateful to Karl Gordon for computing new {\small DIRTY}
models incorporating the LMC2 extinction law, ahead of schedule and
upon our request~; the corresponding data tables are publicly available
on the {\small DIRTY} webpage ({\it http://dirty.as.arizona.edu/}).
This research benefited from the NASA/IPAC Extragalactic Database
(NED), which is operated by the Jet Propulsion Laboratory of Caltech.

\clearpage

\clearpage

\begin{figure}[!ht]
\plotone{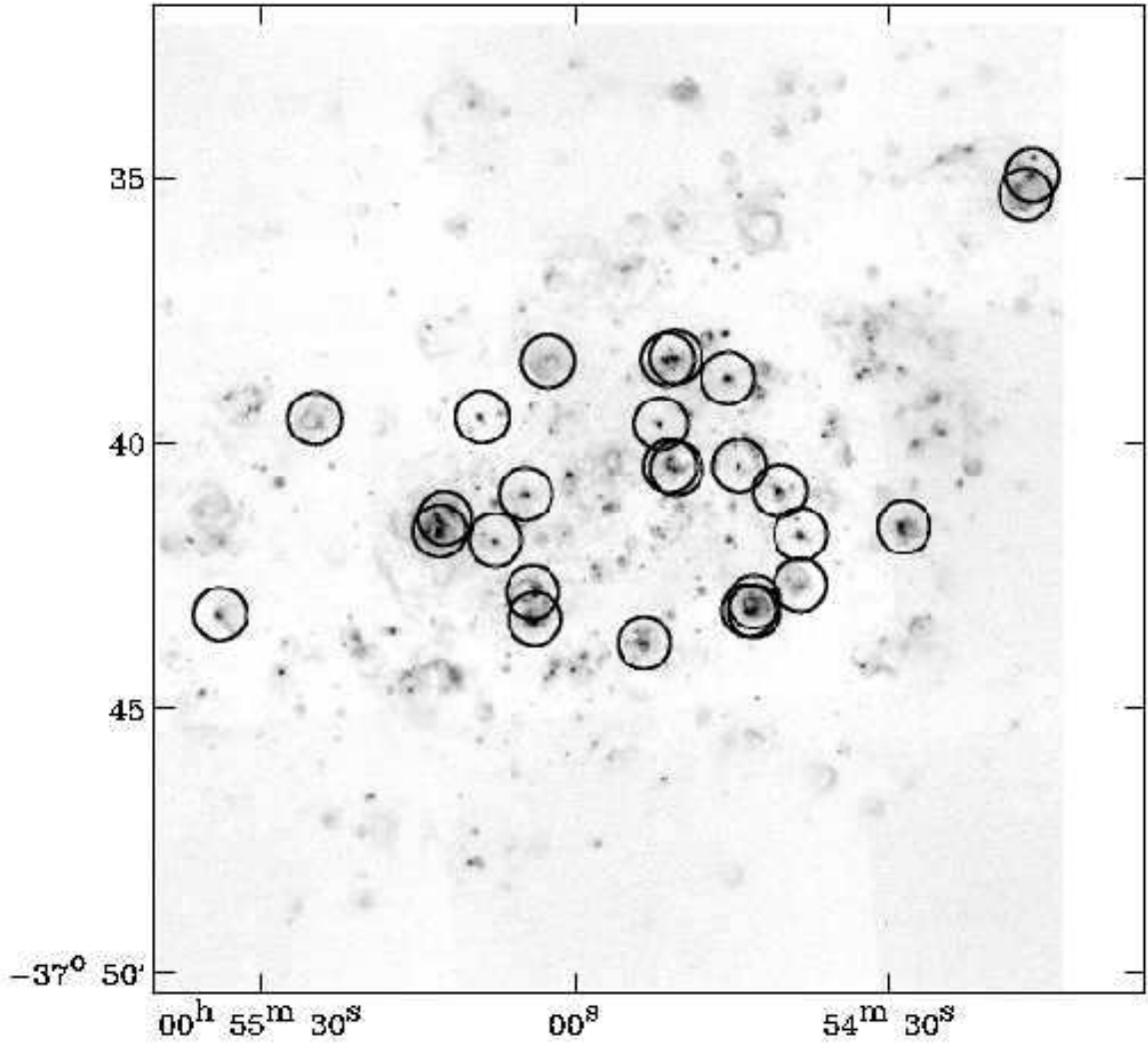}
\caption{H$\alpha$ mosaic, with circles around the \ion{H}{2} complexes
discussed in Sect.~\ref{complexes} superposed. Their diameter is here
arbitrarily fixed at 1\arcmin, for clarity.
The horizontal axis represents the right ascension and the vertical axis the
declination, in the J2000 equatorial coordinate system.
}
\label{fig:halpha}
\end{figure}

\begin{figure}[!ht]
\plotone{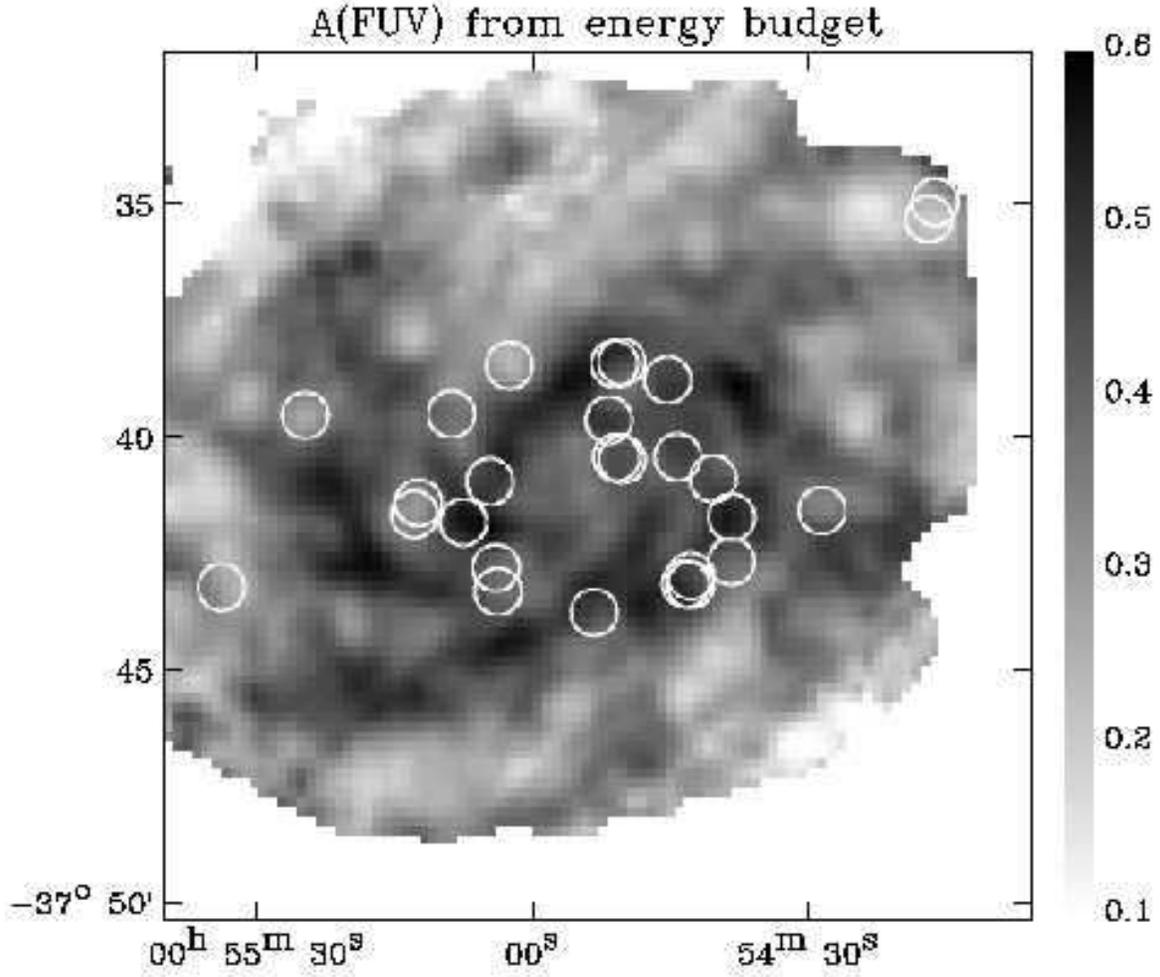}
\caption{Extinction in the FUV band (1516\,\AA) derived from an energy balance
argument: the power absorbed between 0.1 and 3\,$\mu$m is set equal to the total
infrared power. The assumed extinction law is that of \citet{Cardelli},
but results are very similar with the other laws considered in this paper.
The superposed circles represent the locations of the \ion{H}{2} complexes
discussed in Sect.~\ref{complexes}.
The coordinate axes are as in Figure~\ref{fig:halpha}.
}
\label{fig:a_fuv}
\end{figure}

\begin{figure}[!ht]
\plotone{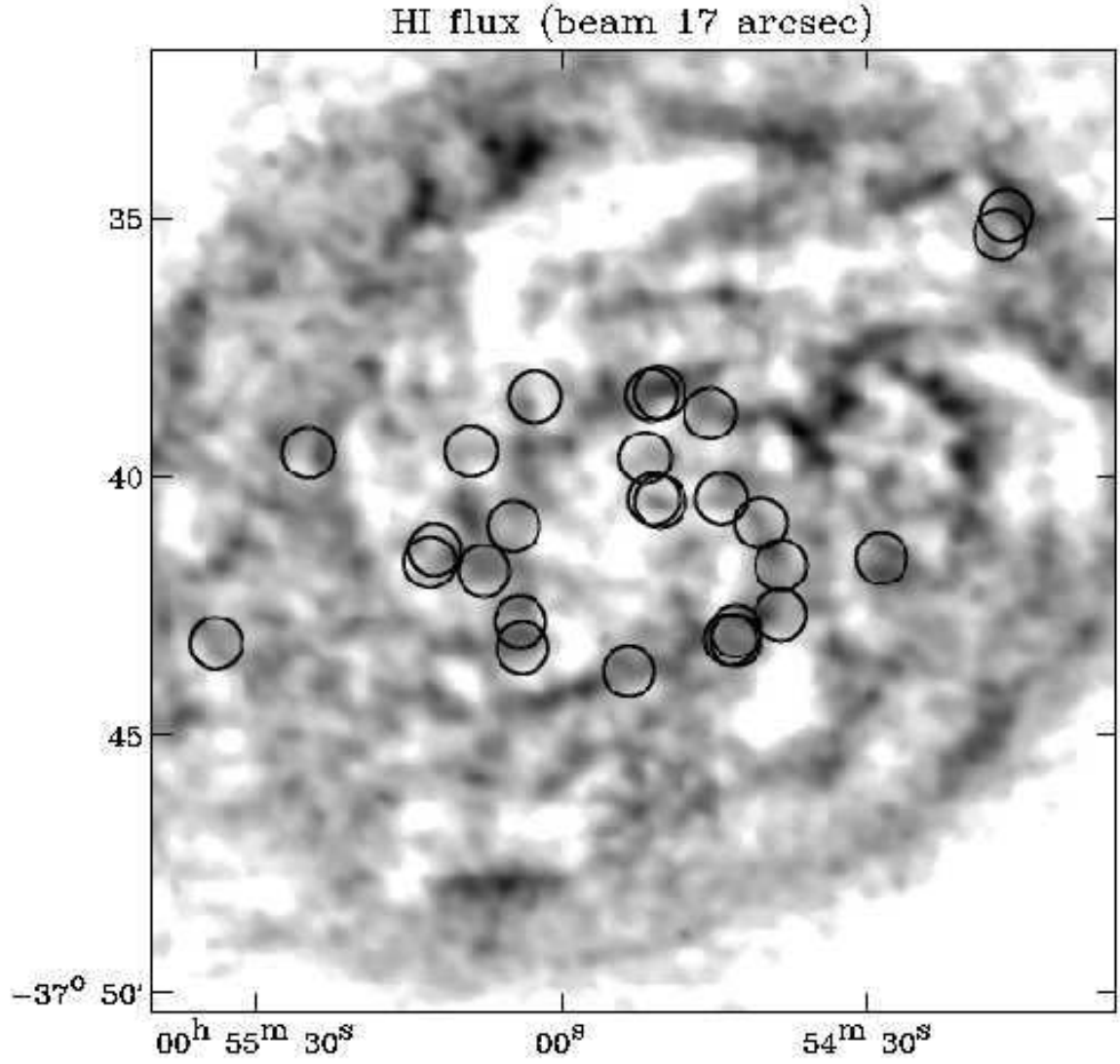}
\caption{\ion{H}{1} flux map from \citet{Puche},
with the same coordinate axes and the
same superposed circles as in Figures~\ref{fig:halpha} and \ref{fig:a_fuv}.
}
\label{fig:hi_apertures}
\end{figure}

\clearpage

\clearpage

\begin{figure}[!ht]
\vspace*{-2cm}
\resizebox{15cm}{!}{\plotone{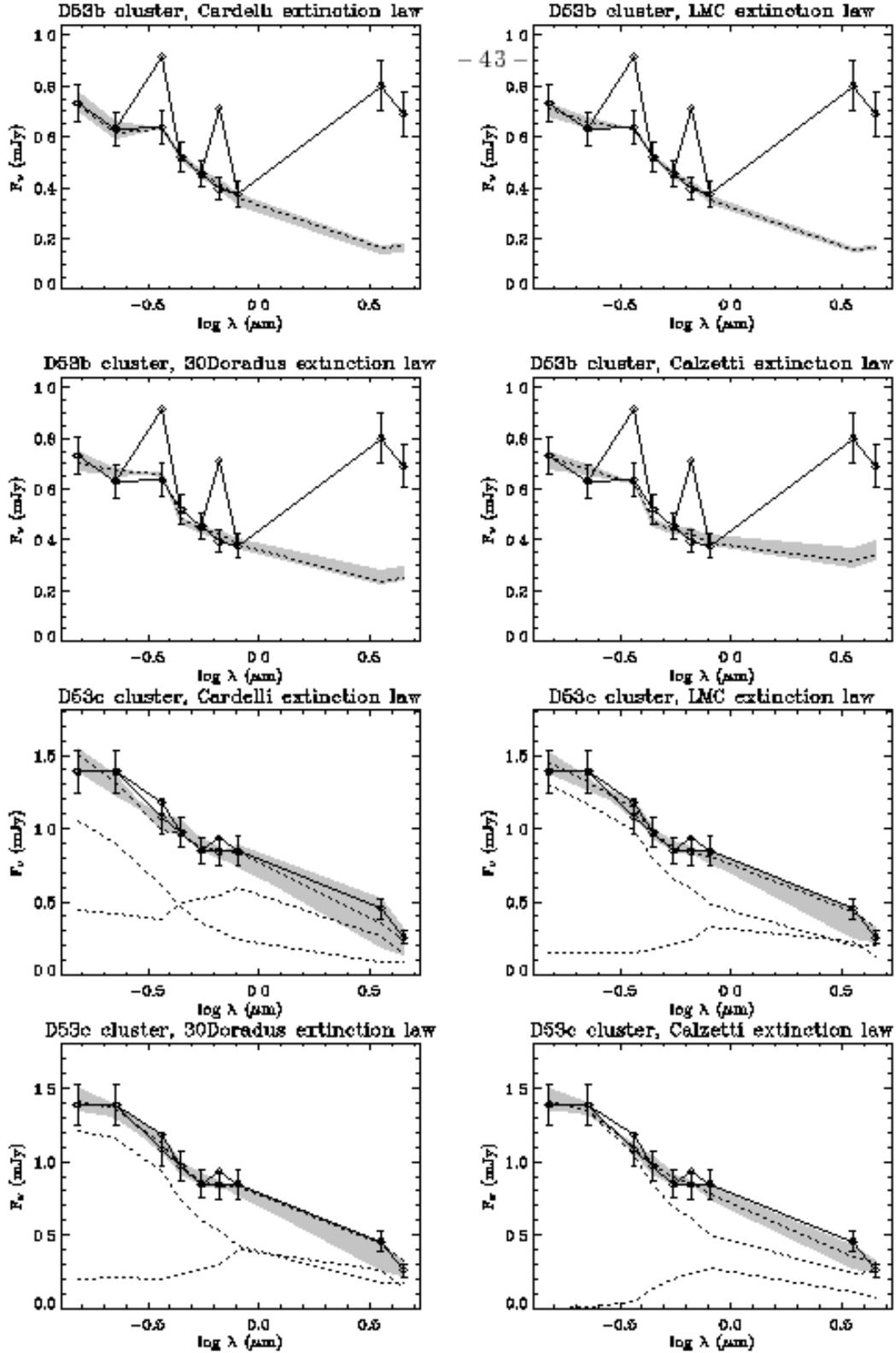}}
\caption{Fits of a few stellar cluster SEDs.
The line connecting diamond symbols with error bars represents the observed
SED after removing the H$\alpha$ and [\ion{O}{2}] contributions from the R
and U bands~; the connected symbols without error bars represent the total
(line $+$ stellar) R and U-band fluxes. The dashed line is the best fit with
the relevant extinction law, and the grey shading shows the $1\sigma$ range
of solutions. When present, two additional dashed lines indicate that two
populations of different ages were fitted and represent their separate SEDs.
}
\label{fig:fits}
\end{figure}

\clearpage

\setcounter{figure}{6}
\begin{figure}[!ht]
\resizebox{15cm}{!}{\plotone{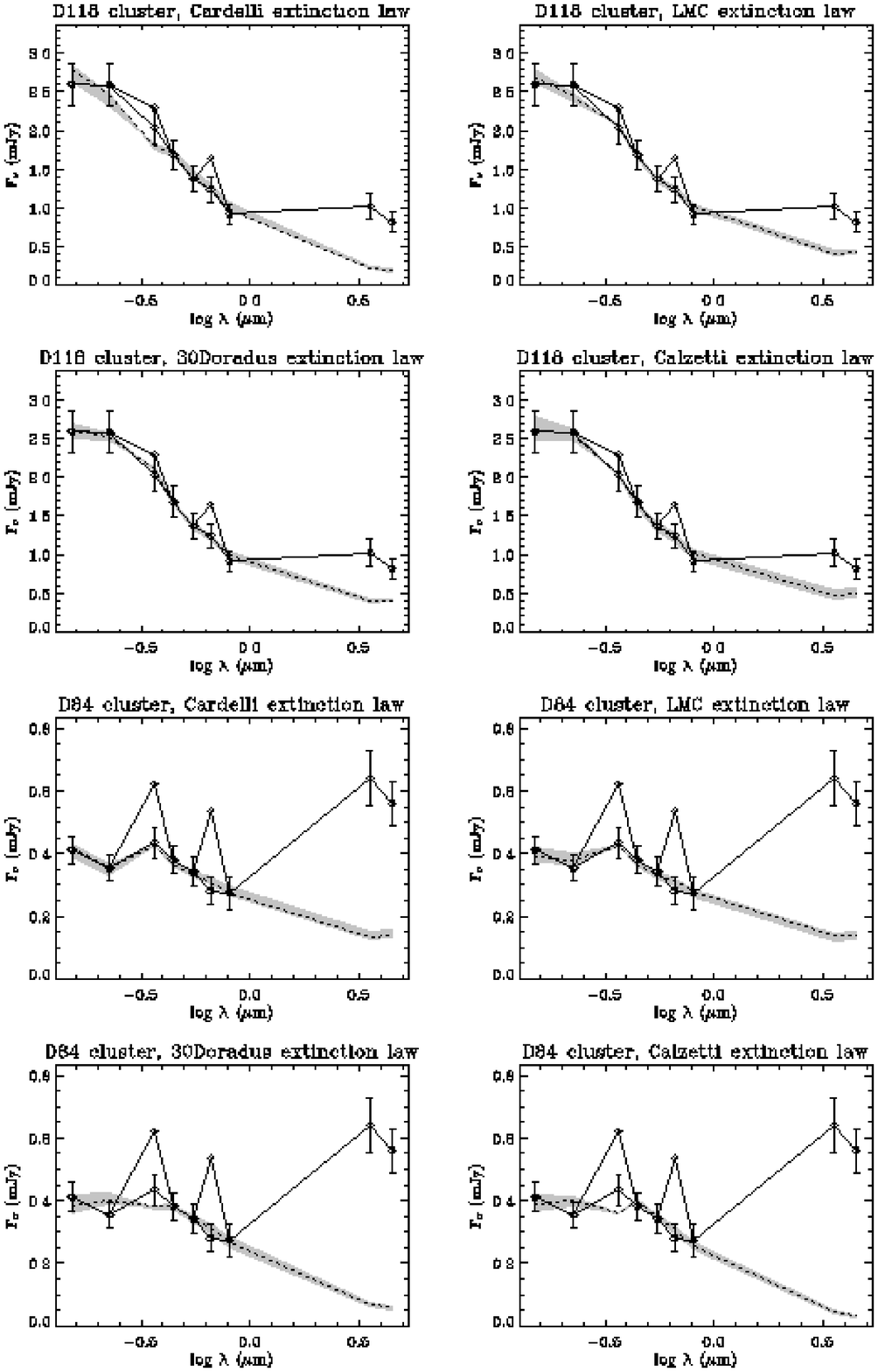}}
\caption{(continued)
}
\label{fig:fits}
\end{figure}

\clearpage

\begin{figure}[!ht]
\vspace*{-2cm}
\resizebox{15cm}{!}{\plotone{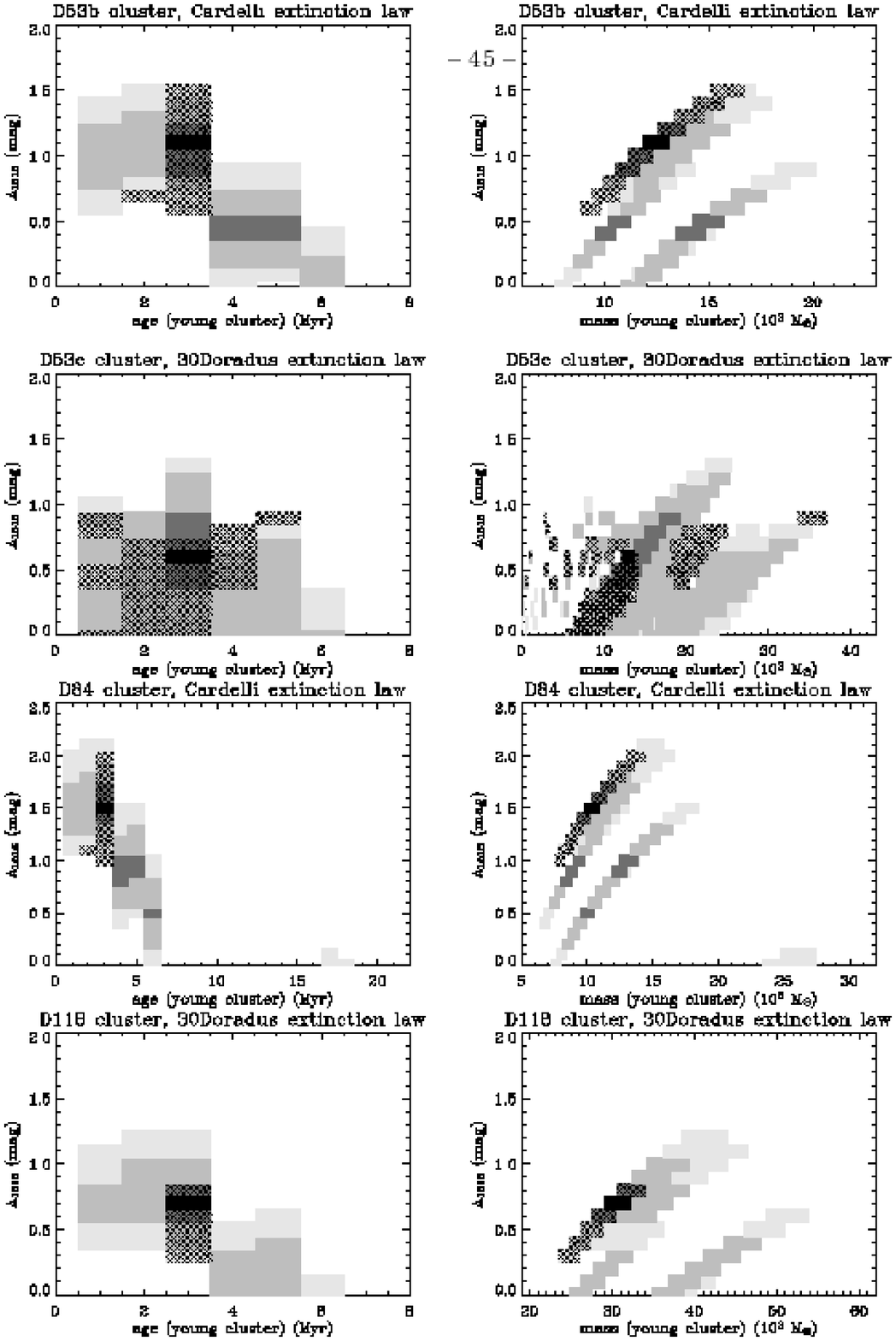}}
\caption{\footnotesize{Confidence interval plots for the fits shown in Figure~\ref{fig:fits}.
The areas of increasing grey intensity represent the $3\sigma$, $2\sigma$ and
$1\sigma$ intervals without any constraint on the ionizing photon flux error
(see text), and the black rectangle indicates the best fit. Due to the
discrete nature of the extinction and age grids, individual solutions with error
intervals ($\pm 0.05$\,mag, $\pm 0.5$\,Myr and $\pm 5$\% for the mass) are plotted
rather than contours. The hatched areas correspond to solutions predicting an
ionizing photon flux within 30\% of the value estimated from the H$\alpha$ and
H$\beta$ maps. No constraint on the FUV/NUV color is implied in $\chi^2$ values.
}}
\label{fig:khi2}
\end{figure}

\clearpage

\begin{figure}[!ht]
\hspace*{-1cm}
\resizebox{9cm}{!}{\rotatebox{90}{\plotone{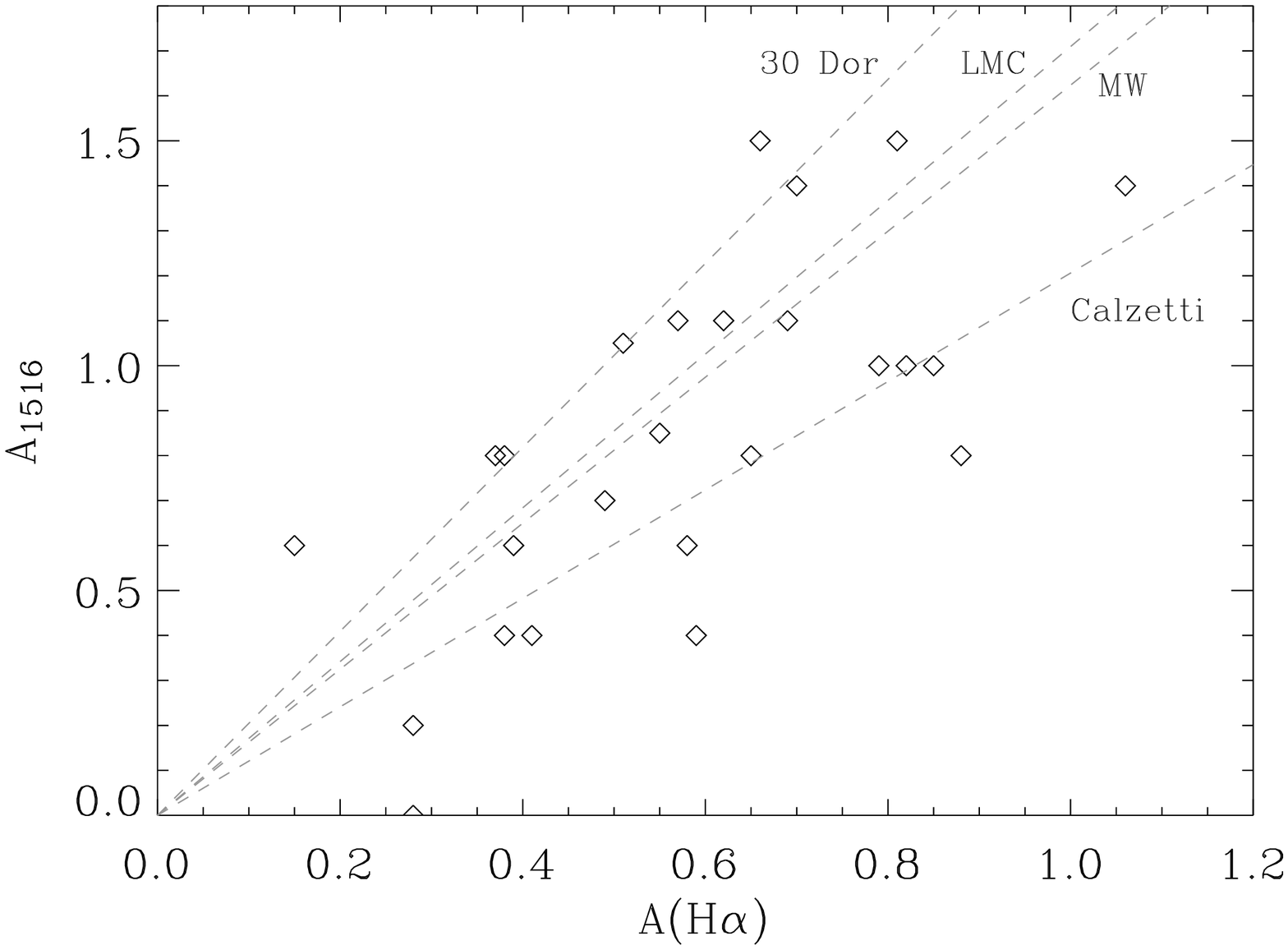}}}
\resizebox{9cm}{!}{\rotatebox{90}{\plotone{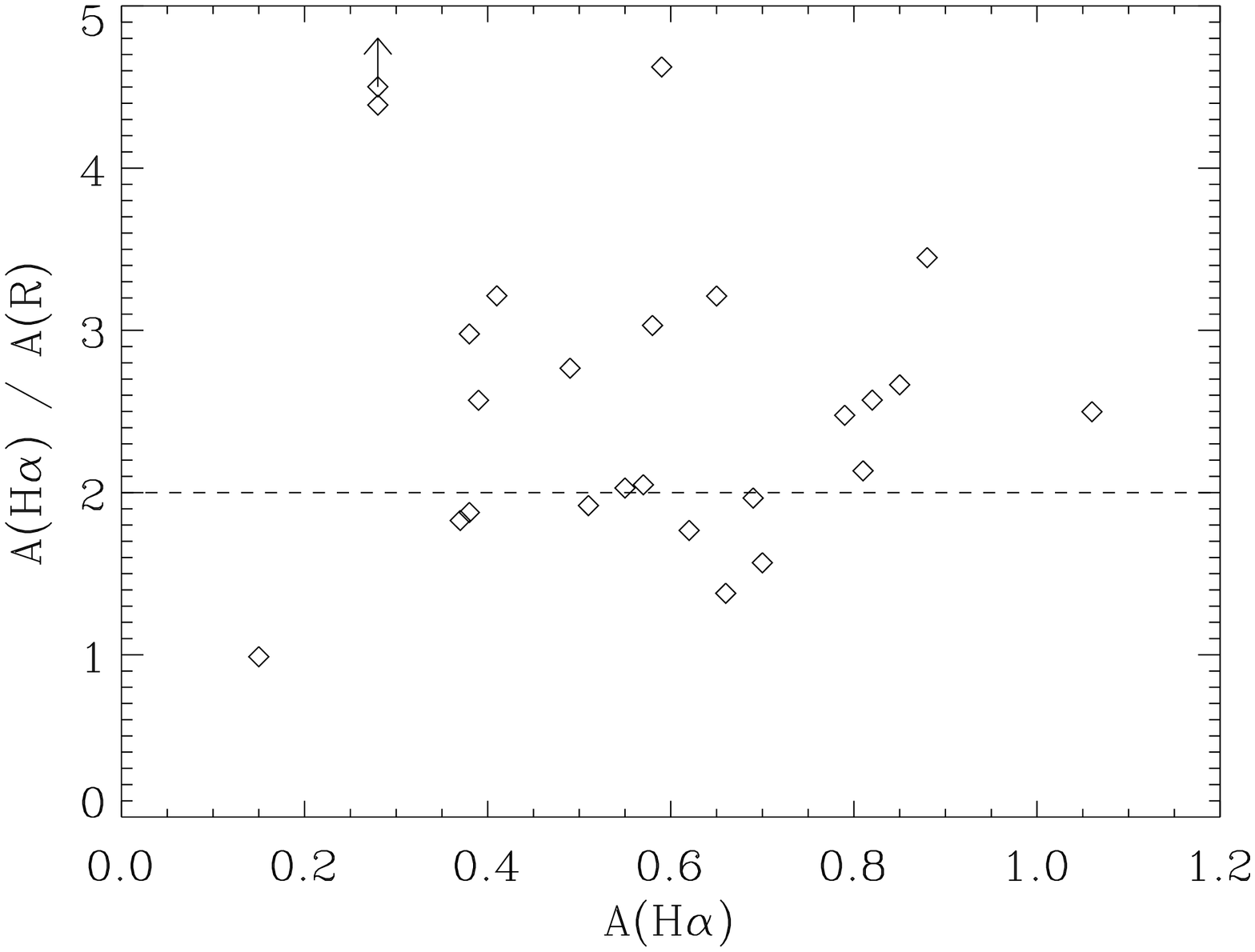}}}
\caption{{\bf Left:} Relationship between the nebular extinction at 6563\,\AA,
estimated from the H$\alpha$/H$\beta$ decrement, and the stellar
extinction at 1516\,\AA\ derived from the SED fits. The dashed lines
represent the prediction of the ratio of stellar to nebular extinction
${\rm A(R)} = 0.5 {\rm A(H}\alpha)$ from \citet{Calzetti94}, extrapolated
to the FUV band using the four extinction laws considered in this paper.
{\bf Right:} Ratio of the nebular extinction at 6563\,\AA\ to the stellar
extinction in the R band, removing the ambiguity on the extinction law
which was appropriately chosen for each cluster, according to the fit
results. The dashed line shows again the prediction from \citet{Calzetti94}.
The two points at ${\rm A(H}\alpha) = 0.28$ and
${\rm A(H}\alpha)/{\rm A(R)} \simeq 4.4$ are D6 and D7-8, which lie outside
the H$\beta$ map and for which we adopted the H$\alpha$/H$\beta$ decrement
of \citet{Webster}~; the fitted stellar extinction of D7-8, being zero,
was replaced by its upper limit (Table~\ref{tab_fits}).
}
\label{fig:a_ha_a1516}
\end{figure}

\begin{figure}[!ht]
\resizebox{12cm}{!}{\rotatebox{90}{\plotone{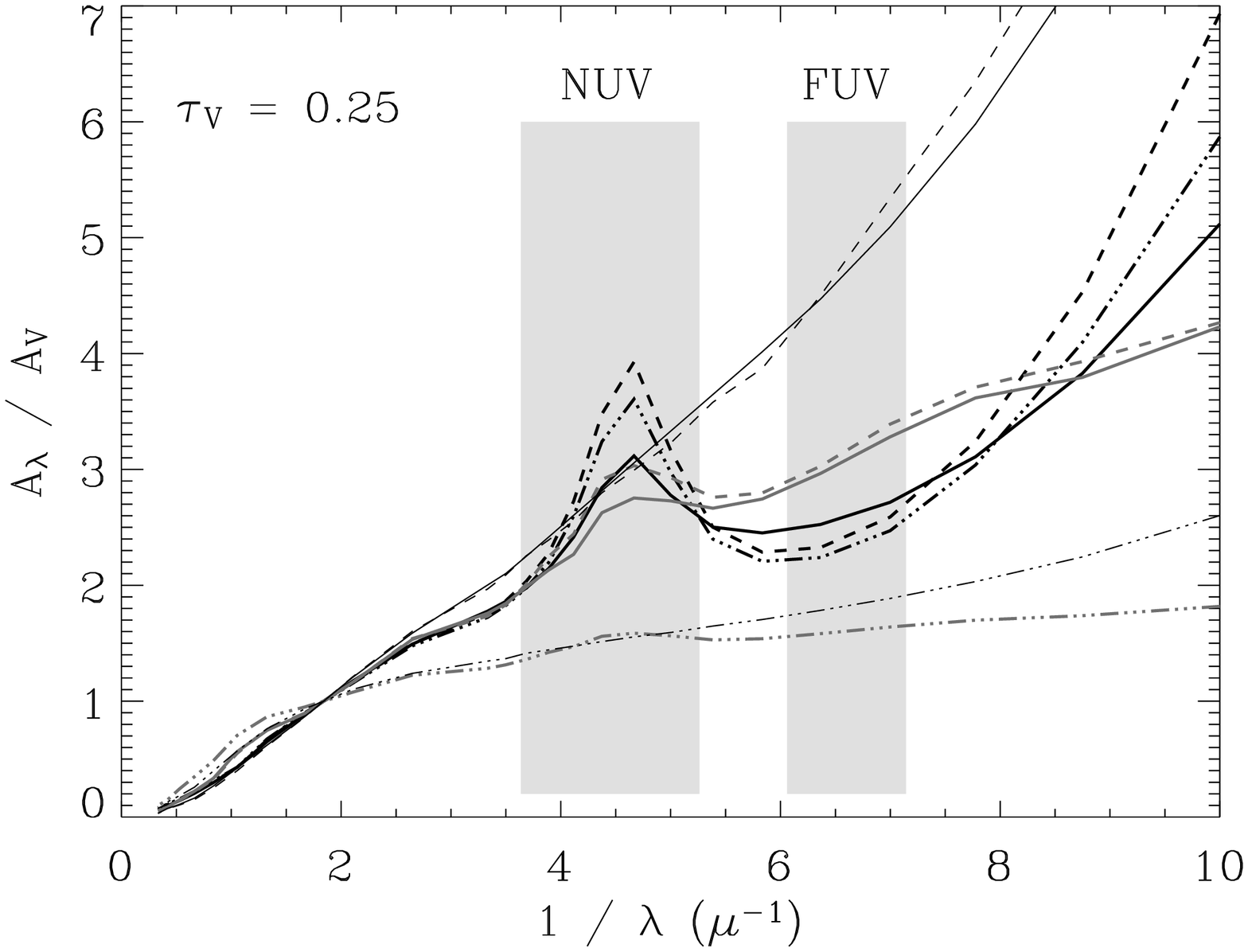}}}
\caption{
Extinction and attenuation laws predicted by the {\small DIRTY} model,
with an opacity $\tau_{\rm V} = 0.25$ (corresponding to $A_{1516} = 0.7$
using the Cardelli extinction law) for the Milky Way (thick black curves),
the LMC2 region (thick grey curves) and the SMC (thin black curves).
The extinction laws are shown as continuous lines~; the attenuation laws
in the homogeneous dust distribution case as dashed lines~; and the
attenuation laws in the clumpy dust distribution case as three-dots-dash
lines. The approximate full widths at half maximum of the Galex transmission
curves are indicated.
}
\label{fig:att_laws}
\end{figure}

\begin{figure}[!ht]
\hspace*{-0.5cm}
\resizebox{18cm}{!}{\rotatebox{90}{\plotone{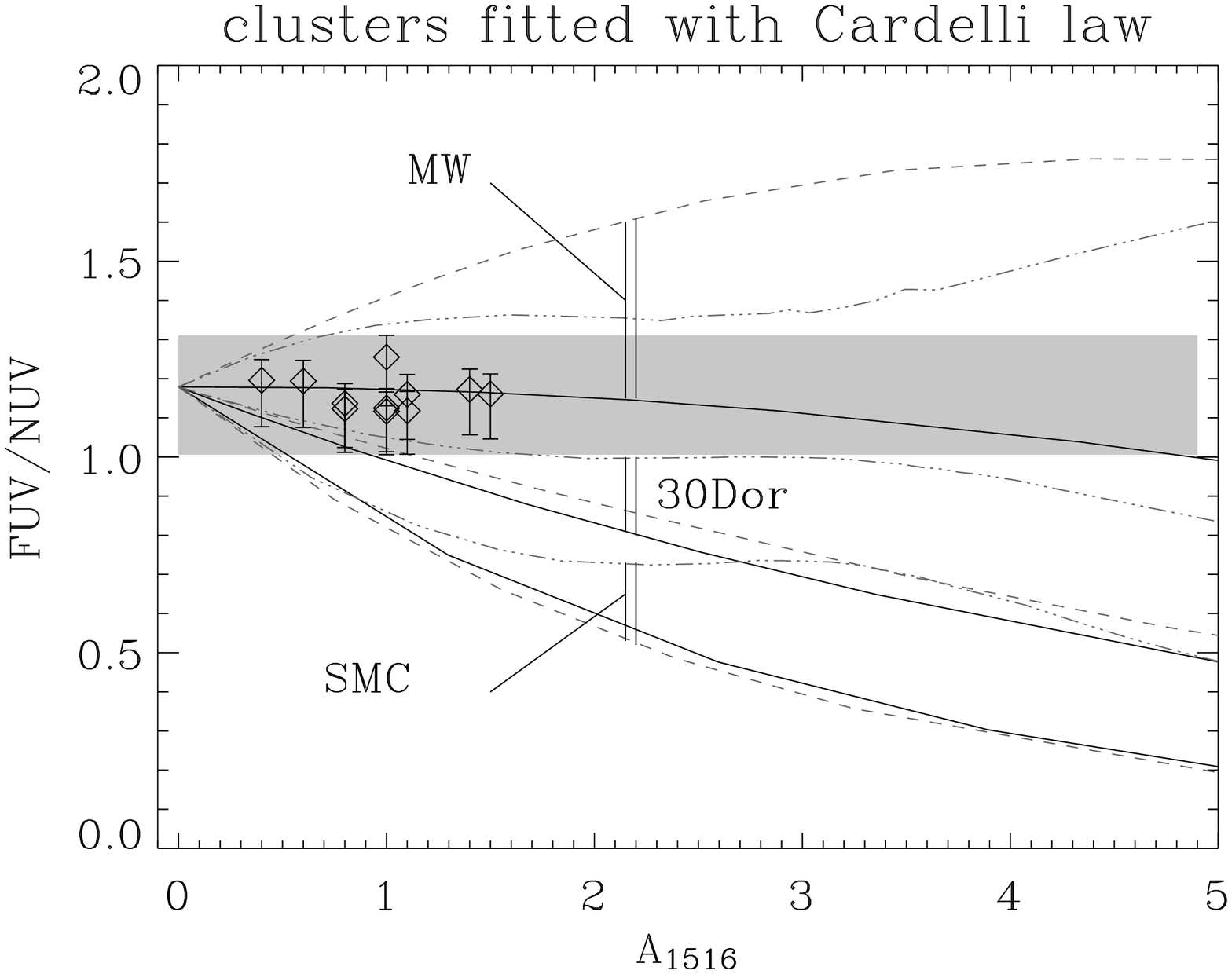}}\rotatebox{90}{\plotone{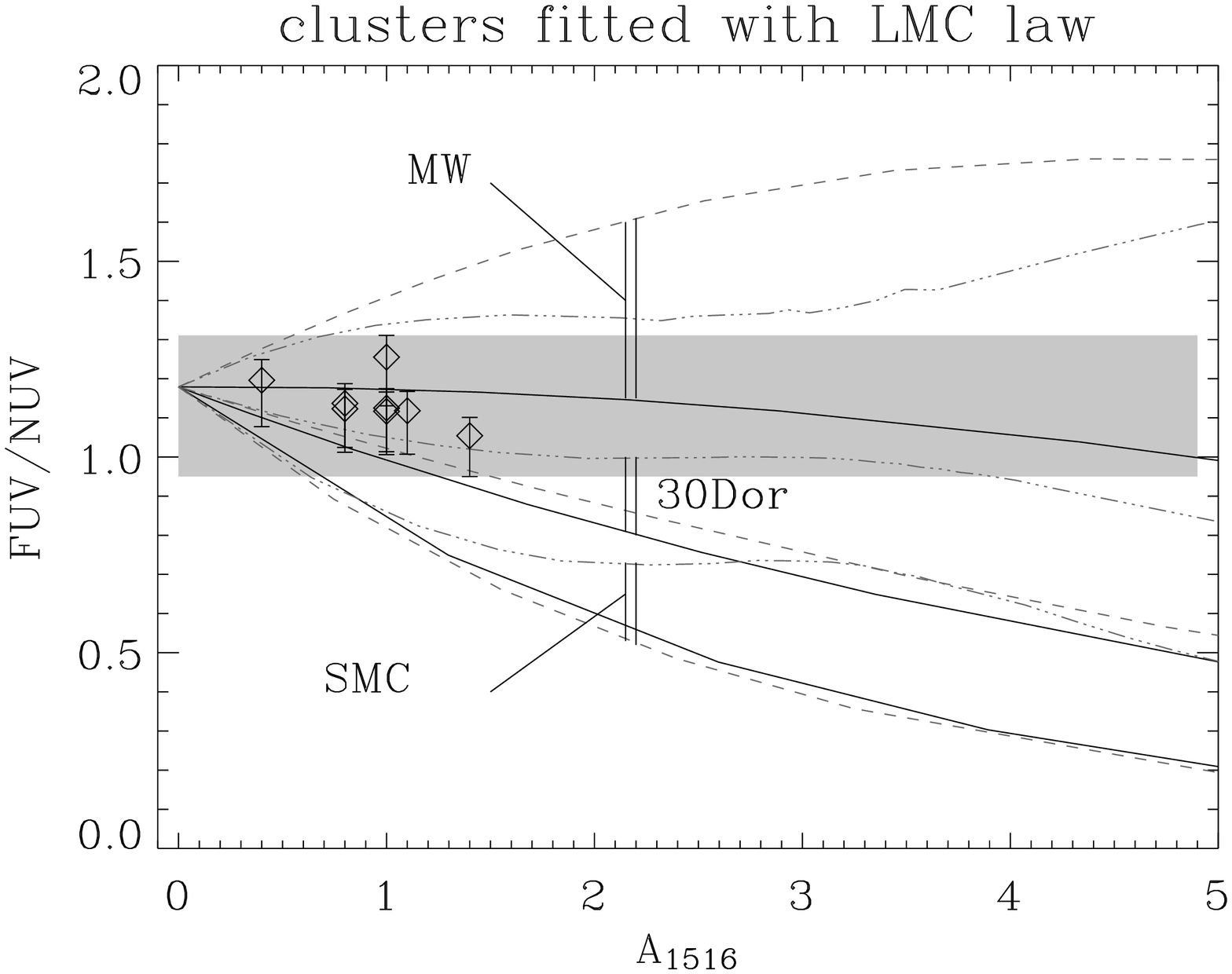}}}
\hspace*{-0.5cm}
\resizebox{9cm}{!}{\rotatebox{90}{\plotone{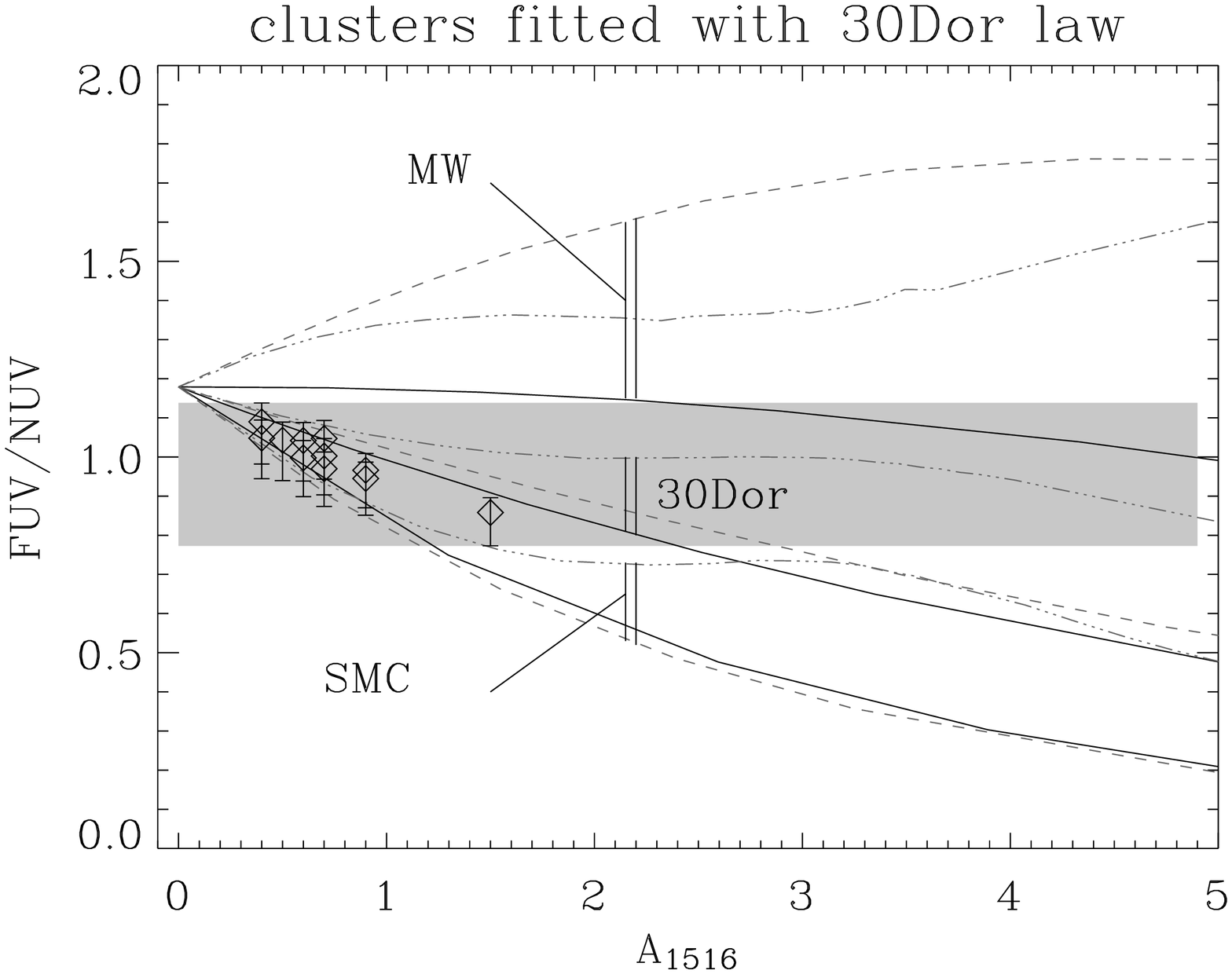}}}
\caption{
Predictions of the {\small DIRTY} model for the observed FUV/NUV
flux density ratio of a 3\,Myr-old cluster, as a function of the extinction
at 1516\,\AA. The continuous lines represent the pure extinction laws
folded in the {\small DIRTY} model: Milky Way, LMC2 (called 30\,Doradus here)
and SMC. The dashed lines represent the corresponding attenuation laws in the
homogeneous case, and the three-dots-dash lines the attenuation laws
in the clumpy case. The vertical double bars show the total range of
FUV/NUV colors derived from each extinction law at $A_{1516} = 2.2$,
for easier identification.
On each panel, the observed FUV/NUV ratios of the clusters well fitted with
each considered extinction law (Cardelli, LMC and 30\,Doradus) are
overplotted. The clusters with unconstrained extinction laws (well fitted
with all three laws) are omitted~; some clusters are in common for the
Cardelli and LMC laws (see Table~\ref{tab_fits}).
The error bars account for the color variations expected when the age of the
clusters is allowed to vary from 1 to 5\,Myr. The maximal range of color
variations, including these error bars, is indicated by a grey rectangle.
}
\label{fig:fuv_nuv_predictions}
\end{figure}

\begin{figure}[!ht]
\resizebox{12cm}{!}{\rotatebox{90}{\plotone{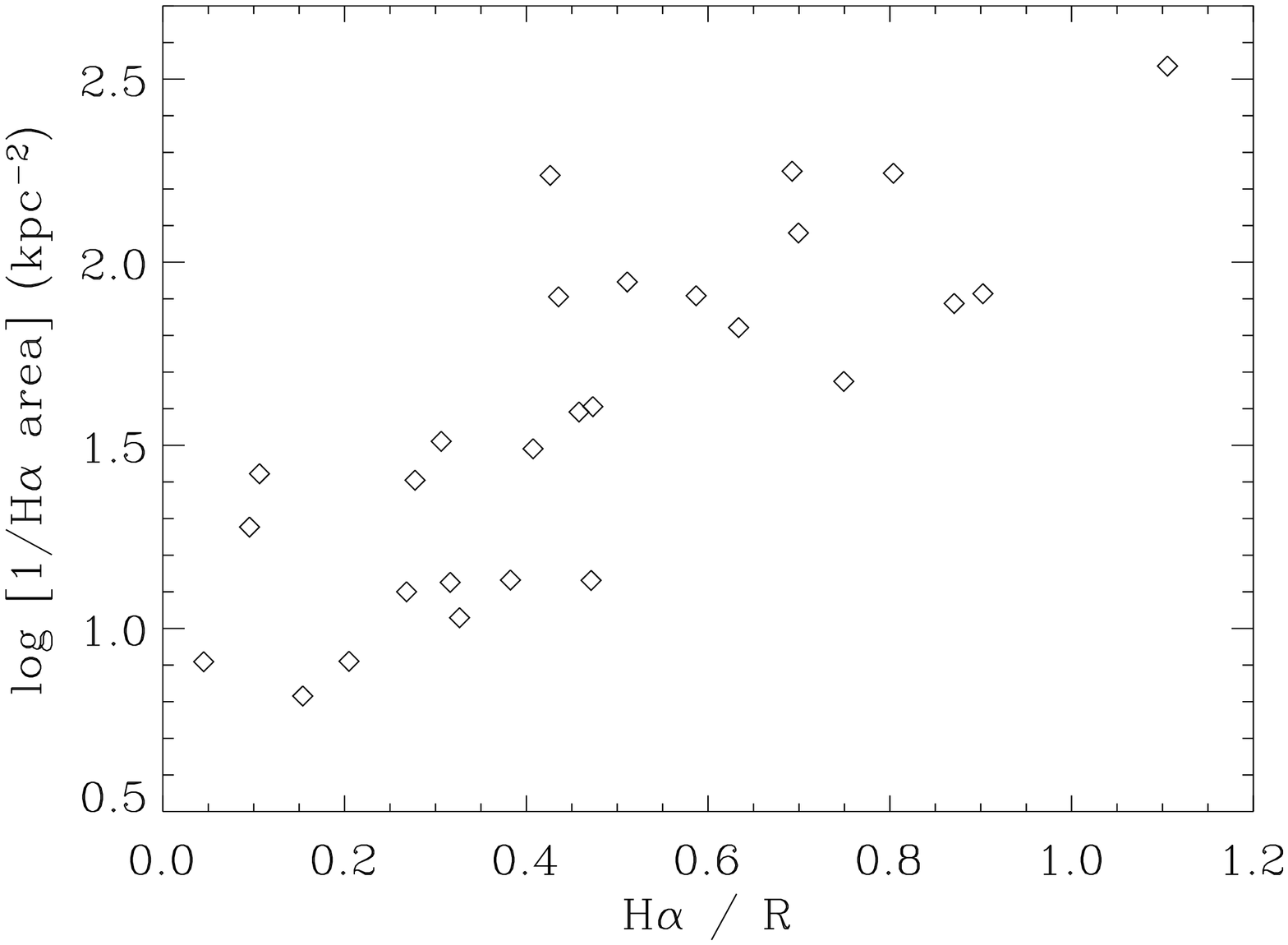}}}
\caption{Relation between the H$\alpha$/R flux ratio,
measured inside the stellar ``aperture'', and the inverse projected
surface area of the entire \ion{H}{2} region. The area was summed inside
an aperture chosen to encompass as much of the \ion{H}{2} region as
possible without including nearby \ion{H}{2} regions, and using only
the pixels above $6 \sigma$.
}
\label{fig:ew_compactness}
\end{figure}

\begin{figure}[!ht]
\resizebox{12cm}{!}{\rotatebox{90}{\plotone{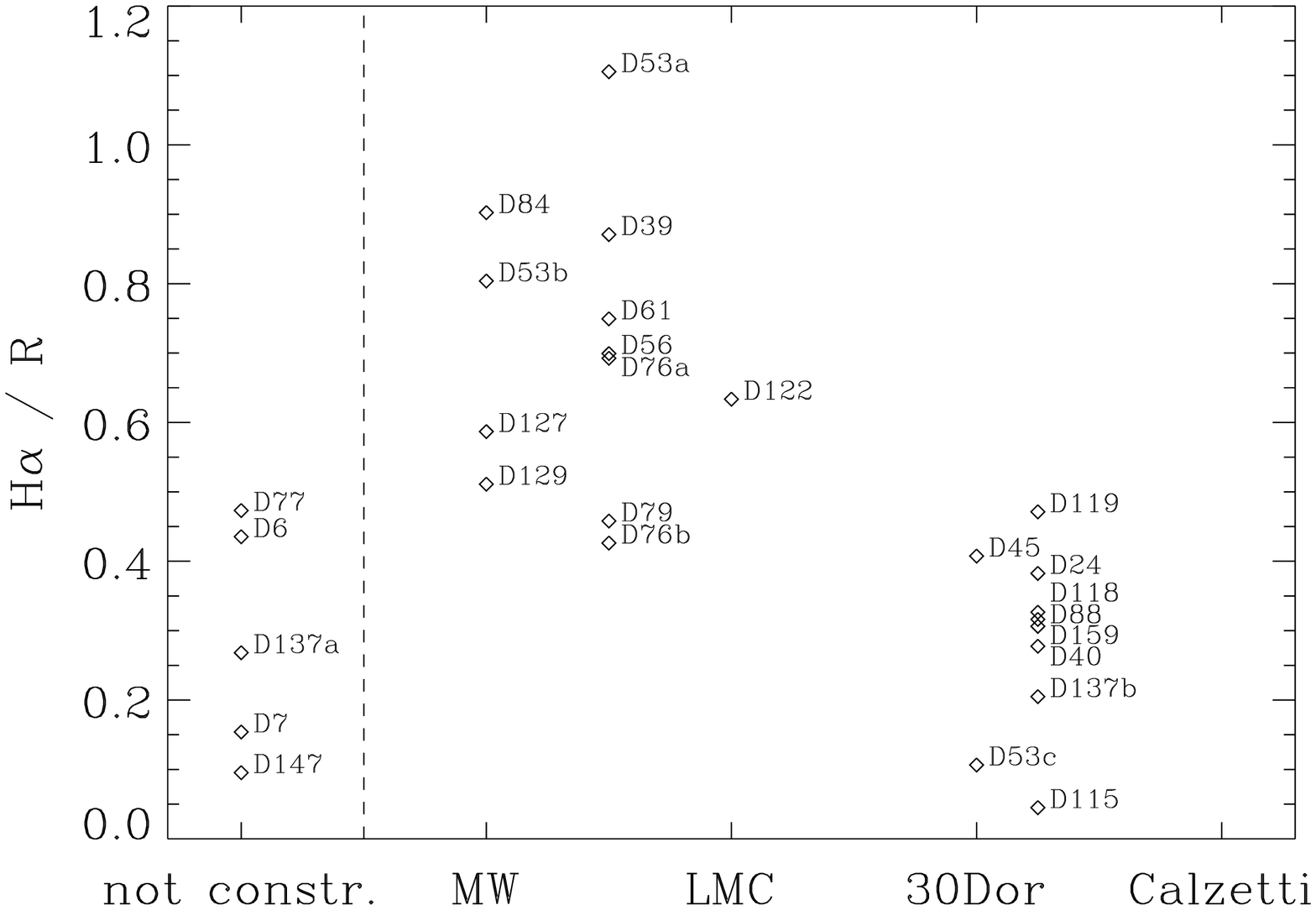}}}
\caption{Compactness of the regions (as quantified by the H$\alpha$/R
flux ratio) and extinction laws that are able to reproduce the observed spectral
energy distributions from the FUV to the I band (see text). The label
``not constr.'' means that the extinction law shape is not constrained
by the stellar SED fits.
}
\label{fig:extlaw_ew}
\end{figure}

\begin{figure}[!ht]
\resizebox{12cm}{!}{\rotatebox{90}{\plotone{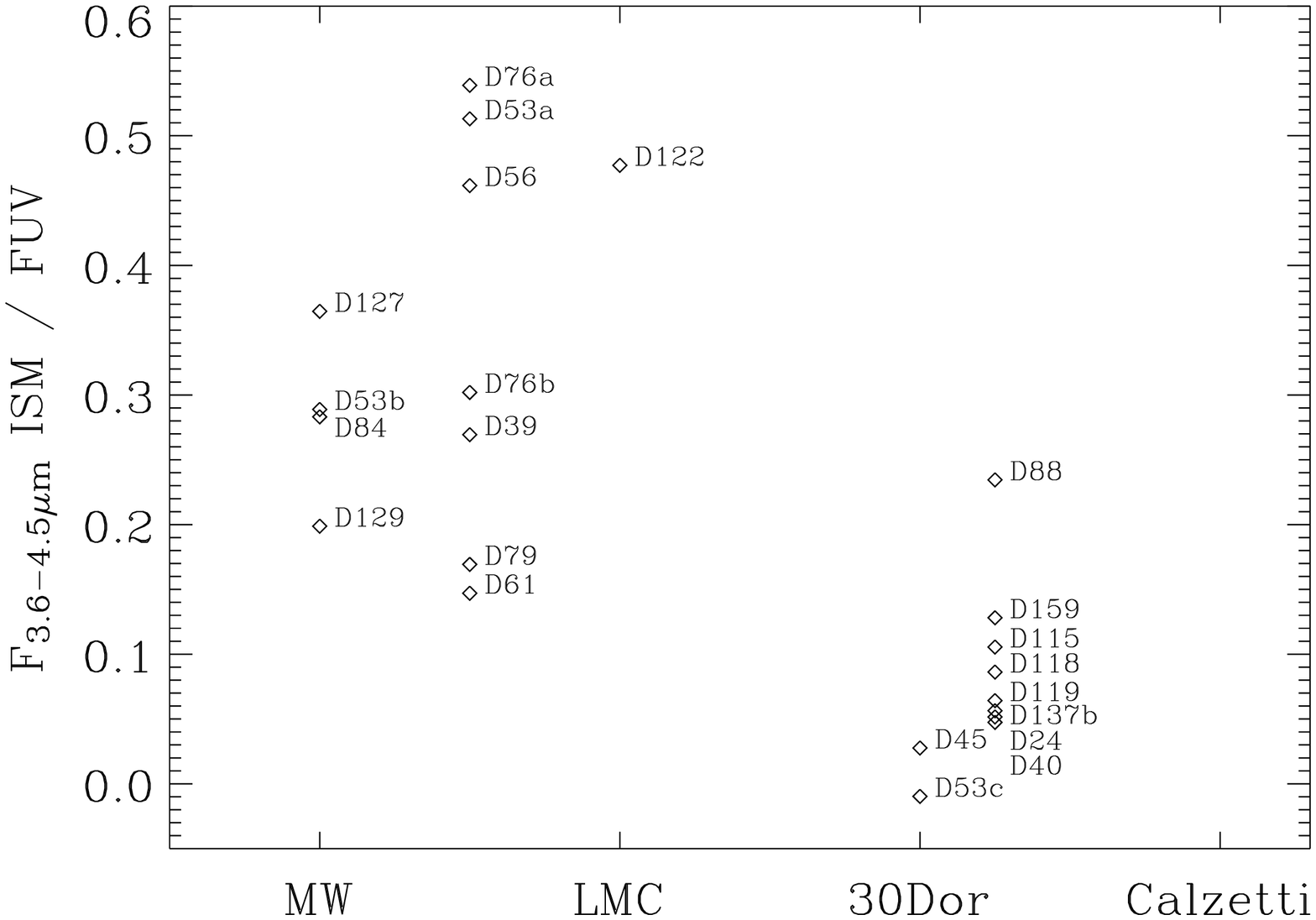}}}
\caption{Ratio of the estimated interstellar component flux at 3.6-4.5\,$\mu$m
to the extinction-corrected FUV flux (chosen as normalization because
it should be closely correlated with the heating intensity of the considered
dust species), and applicable extinction laws.
}
\label{fig:extlaw_ism}
\end{figure}

\begin{figure}[!ht]
\resizebox{12cm}{!}{\rotatebox{90}{\plotone{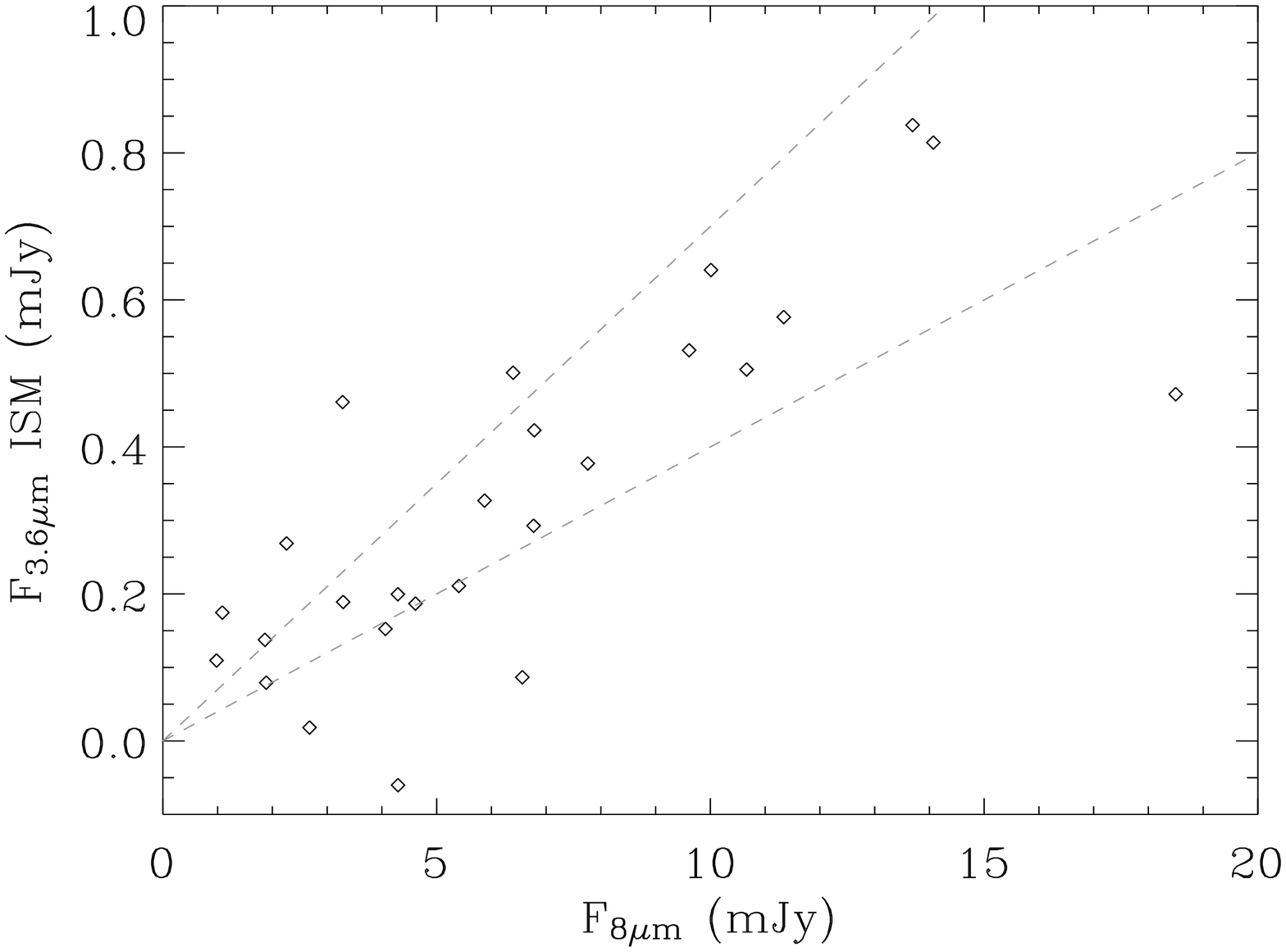}}}
\caption{Correlation between the interstellar component at 3.6\,$\mu$m and
the aromatic band emission at 8\,$\mu$m, measured within the stellar apertures,
at the 8\,$\mu$m angular resolution. The dashed lines show brackets on the
interstellar $F_{3.6}/F_8$ flux ratio derived by \citet{Lu}.
The relation with the 24\,$\mu$m emission is apparently less tight,
but we are severely limited by the degraded angular resolution at
this wavelength, about twice as large as at 8\,$\mu$m.
}
\label{fig:f8_f3}
\end{figure}

\begin{figure}[!ht]
\resizebox{15cm}{!}{\plotone{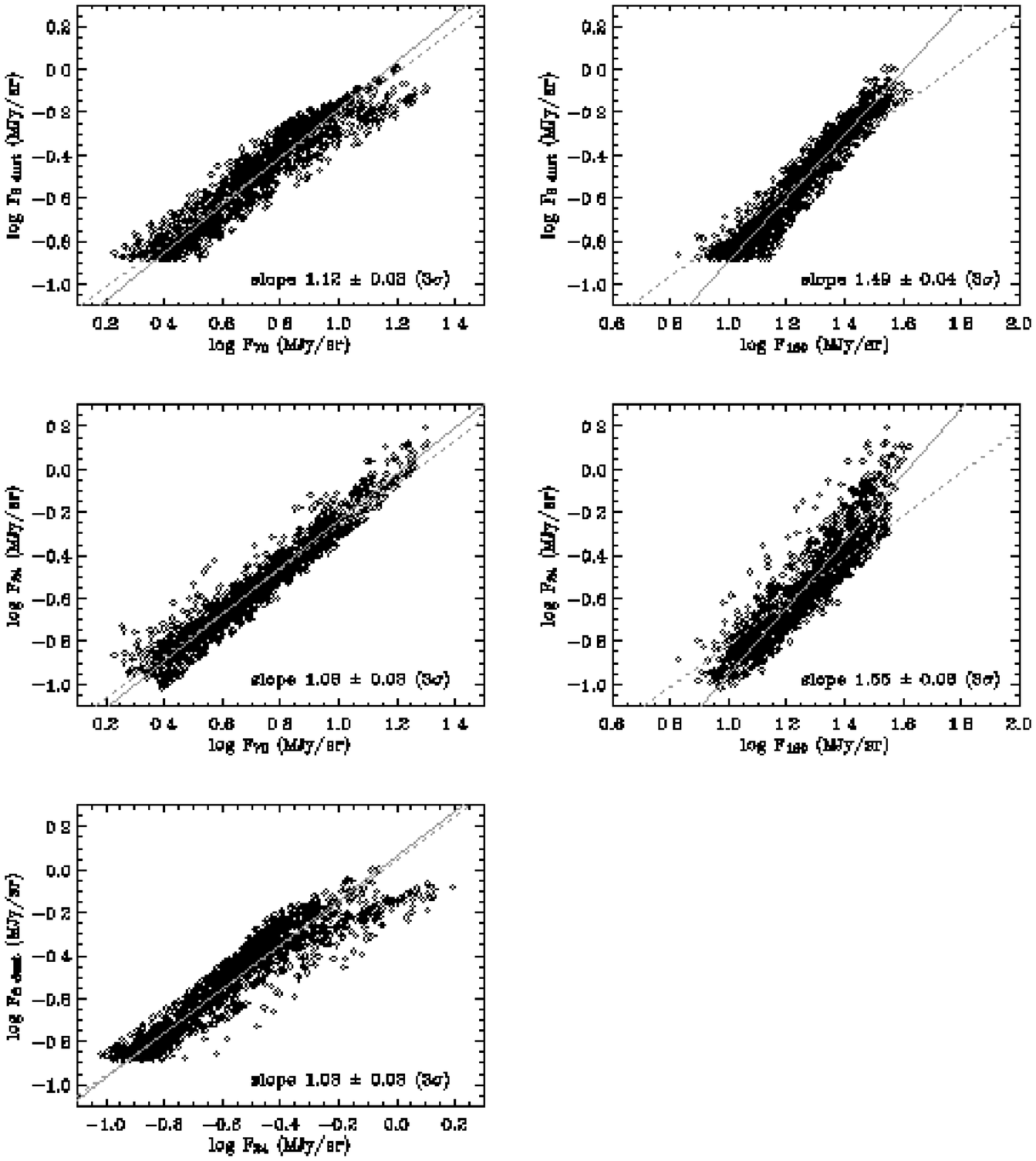}}
\caption{Correlation between the emission at 8 and 24\,$\mu$m and the
emission at 70 and 160\,$\mu$m, at the angular resolution of the
160\,$\mu$m image. The solid lines indicate the least absolute deviation fits,
and the dashed lines have a slope of 1.
}
\label{fig:pah_bg}
\end{figure}

\end{document}